\documentclass[conference]{IEEEtran}

\pdfoutput=1

\let\labelindent\relax

\usepackage{amsmath}
\usepackage{bbding}
\usepackage{pifont}
\usepackage{xfrac}
\usepackage{graphicx}
\usepackage[font=footnotesize,caption=false]{subfig}
\usepackage{stmaryrd}
\usepackage{array}
\usepackage{enumitem}
\usepackage{mdwmath}
\usepackage{mdwtab}
\usepackage{eqparbox}
\usepackage{slashbox}
\usepackage{stfloats}
\usepackage{moreverb}
\usepackage{listings}
\usepackage{algpseudocode}
\usepackage{algorithm}
\usepackage[hyphens]{url}
\usepackage{color}
\usepackage{multirow}
\usepackage{setspace}
\usepackage[utf8]{inputenc}
\usepackage[T1]{fontenc}
\usepackage{microtype}
\usepackage{flushend}
\usepackage{datetime}
\usepackage{epsfig}
\usepackage{wrapfig,lipsum,booktabs}
\usepackage{url}
\usepackage[hidelinks]{hyperref}

\usepackage{hhline}

\usepackage{xspace}

\usepackage{ifpdf}

\definecolor{dkgreen}{rgb}{0,0.7,0.5}
\definecolor{gray}{rgb}{0.5,0.5,0.5}
\definecolor{mauve}{rgb}{0.6,0,0.8}

\newtheorem{definition}{Definition}

\newcommand{\aaf}{\vspace*{-6pt}}
\newcommand{\af}{\vspace*{-3pt}}

\newcommand{\aeg}{\textsc{Centaur}\xspace}

\newcommand{\latin}[1]{{\it #1}}

\IEEEoverridecommandlockouts

\begin{document}






%


\title{Context-aware System Service Call-oriented Symbolic Execution of Android Framework \\ with Application to Exploit Generation}

\author{\IEEEauthorblockN{Lannan Luo$^{\dag \ast}$, Qiang Zeng$^{\ddag \ast}$, Chen Cao$^\S$, Kai Chen$^\S$, Jian Liu$^\S$, \\ Limin Liu$^\S$, Neng Gao$^\S$, Min Yang$^\wr$, Xinyu Xing$^\dag$, and Peng Liu$^\dag$}
\IEEEauthorblockA{$^\dag$The Pennsylvania State University, University Park, PA, USA\\
$^\ddag$Temple University, PA, USA\\
$^\S$Institute of Information Engineering, Chinese Academy of Sciences, China\\
$^\wr$Fudan University, China\\
\{lzl144, xxing, pliu\}@ist.psu.edu, \{qzeng\}@temple.edu, \\ \{caochen, chenkai, liujian6, liulimin, gaoneng\}@iie.ac.cn, \{m\_yang\}@fudan.edu.cn}
\thanks{$^{\ast}$These two authors have contributed equally.}}


\maketitle


\begin{abstract}
Android Framework is a layer of software that exists in every Android system
managing resources of all Android apps.
A vulnerability in Android Framework can lead to severe hacks,
such as destroying user data and leaking private information.
With tens of millions of Android devices unpatched due to Android fragmentation,
vulnerabilities in Android Framework certainly attract attackers to exploit them.
So far, enormous manual effort is needed to
craft such exploits. To our knowledge, no research has been done 
on automatic generation of exploits that take advantage of Android Framework
vulnerabilities. We make a first step towards this goal by
applying symbolic execution of Android Framework to finding bugs
and generating exploits. 
Several challenges have been raised by the task. 
(1) The information of an app flows to Android Framework
in multiple intricate steps, making it difficult to identify symbolic inputs.  
(2) Android Framework has a complex initialization phase,
which exacerbates the state space explosion problem. 
(3) A straightforward design that builds the symbolic executor as a layer inside the
Android system will not work well: not only does the implementation 
have to ensure the compatibility with the Android system, 
but it needs to be maintained whenever Android gets updated.
We present novel ideas and techniques to resolve the challenges, and have
built the first system for symbolic execution of Android Framework. 
It fundamentally changes the state of the art in exploit generation
on the Android system, and has been applied to constructing
new techniques for finding vulnerabilities. 

\end{abstract}


\section{Introduction}

Android Framework (aka, Android Application Framework) 
contains a set of system services, managing system resources and life cycles of applications~\cite{framework}.
Recently, many vulnerabilities in Android Framework have been 
identified~\cite{CVE-2015-6628, CVE-2016-2496, CVE-2016-3759, CVE-2016-3750}.
Vulnerabilities in Android Framework can cause severe security consequences; e.g.,
malicious apps can exploit them to steal user passwords, take pictures in the background, 
launch UI spoofing attacks, and tamper with user data~\cite{task,kratos,stagefright}.
On the other hand, due to Android fragmentation among the 1.4 billion active devices~\cite{users}, 
tens of millions of Android devices are left unpatched, ``turning devices into a toxic 
hellstew of vulnerabilities''~\cite{toxic}. 
Given the severe security consequences and the large number of vulnerable Android devices,
attackers are certainly motivated to exploit Android Framework vulnerabilities.

So far, such exploits have been mainly crafted manually;\footnote{Exceptions exist; e.g., fuzzing 
has been used for revealing input validation bugs in Android 
Framework~\cite{mulliner2009injecting, inputvalid}; the bug-revealing inputs
can be used to launch DoS attacks trivially. We consider general types of vulnerabilities.}
attackers need to go through the complex logic of Android Framework
to figure out the exploit,
which is a challenging, laborious, and lengthy process.
Previous researches have shown the feasibility of automatic exploit generation,
but mostly deem Windows or Unix stand-alone 
executables to be victims~\cite{apeg, aeg}. (Certainly, 
once an exploit succeeds, e.g., in hijacking the
control flow of the victim executable, the \emph{payload} may further target other victims;
how to construct a payload is beyond the scope of this paper.)


There has been no previous work that automatically generates exploits taking
advantage of Android Framework vulnerabilities. Such an exploit is embedded 
inside a malicious app and may target another app. The exploit generation thus has to consider 
multiple entities: the malicious app, the system services in Android Framework,
and the victim app. For example, in order to launch a task hijacking attack~\cite{task}, which seeks to
place a malicious activity in the back stack where the victim activity resides,  
the malicious app invokes the Activity Manager Service, which further communicates with
a set of other system services and finally adds the malicious activity to the back stack hosting
the victim activity.\footnote{The malicious activity can then be used to
conduct, e.g., UI spoofing.}
A new family of Android Framework specific problems
have to be investigated and dealt with, e.g., how to handle the interaction between system services, 
how the states of apps are represented in Android Framework.
Therefore, it will not work out by porting an existing exploit generation system that
targets stand-alone executables to Android for exploit generation. 
  
In addition, existing techniques typically generate exploits
as some simple form of inputs of stand-alone executables, 
such as a command line argument, a format string, a network packet,
a piece of file metadata, etc. 
In contrast, the exploit we consider here is part of a malicious app,
and comprises the malicious app's configuration (i.e., the 
manifest file) and code that issues system service calls. 
This is another reason why existing exploit generation techniques 
are not applicable here.




This work aims at automatic generation of exploits that take advantage of Android 
Framework vulnerabilities. It has multiple security implications. First, it advances the
state of the art in exploit generation on the Android system, and
upgrades attackers' capabilities of crafting zero-day exploits. 
With the large number of automatically generated exploits, 
even after an exploit is well known and suppressed, 
it will be trivial for attackers to roll out fresh exploits. 
This calls for a better understanding of attackers' capabilities and
more powerful defense against such exploits.
Second, it can be applied to defense systems. For instance, the generated exploits can be fed into automated 
malware signature generation algorithms by defenders without seeing real-life malware~\cite{aeg,Costa:2007:bouncer}.
Third, as shown in our evaluation, the techniques proposed in this work can be used
to find Android Framework vulnerabilities.  

An Android Framework vulnerability is exploited usually because some security property $P_s$ is violated. 
For example, a vulnerability due to insufficient permission checking
is exploited if the property that ``\emph{the target resource can only be
accessed with proper permission}'' is violated; a task hijacking attack succeeds when the security 
property that ``\emph{the malicious activity should not be placed onto the back stack
hosting the victim activity}'' is broken. Thus, among all possible execution 
paths (of Android Framework),
we apply symbolic execution to searching for paths where a given $P_s$ can be violated;
if such a path exists, a suspected vulnerability is found and the corresponding
path condition may be used to construct exploits. The exploits then can be used to validate
the suspected vulnerability. 


\noindent \textbf{\emph{Challenges} \enspace} Several challenges are
raised by symbolic execution of Android Framework for exploit generation.


First, Android Framework hosts a set of system services serving
all apps; thus, it has a large number of 
data structures that store the information of the different entities 
(system services and apps), e.g., a list 
that stores the permissions granted to the installed apps, and stacks that store
activities of different apps. From the perspective of exploit generation, 
a vital step is to correctly identify which variables are derived 
from the malicious app and specify them as symbolic inputs (since this means 
the variables may be manipulable by attackers), such that all 
possible values of these variables can be considered by path exploration.
However, 
it is challenging to determine, among the
numerous variables in Android Framework, which have been
derived from the malicious app. Previous work~\cite{taintdroid,flowdroid}
applies tainting to revealing whether taints originated from specific sources (e.g., 
return values of some system calls in the app) may propagate to specific sinks. 
But no previous work has been done for comprehensively tracking the information flow from a whole 
app to the underlying system services. Such tracking is very difficult, if not impossible,
as the information spreads throughout app installation, system service initialization, and 
starting the app. 
Considering the complexity of these steps, 
a precise tainting is very hard to achieve, as it requires enormous work for
handling overtainting and undertainting~\cite{all-taint}. 

This is a unique challenge, because, as aforementioned, previous exploit generation techniques 
usually create exploits as a simple form of inputs,
and the variables corresponding to the inputs, such as a command line argument and a network packet, 
are easy to identify. But the exploit we consider here
is part of an app comprising configuration and code, and the information flow from
the app to Android Framework is complex. This renders the identification of symbolic inputs difficult.


Second, as Android Framework has a very complex initialization phase, 
the whole-program symbolic execution that starts from the \texttt{main}
function of Android Framework can lead to severe state space explosion.
An alternative to the whole-program symbolic execution is \emph{under-constrained}
symbolic execution~\cite{ucklee,under1,under2}, which can start from an arbitrary function
within the program and thus allows previously-unreachable code to be checked.
Nevertheless, as it skips the initialization phase, the execution context (e.g., the 
type and value information of variables) that could have
been provided by the initialization is missing. For example, consider a virtual 
function call \texttt{r.foo}, where \texttt{r} is a reference of an interface or 
abstract class type; as the real type of the object pointed to by
\texttt{r} is unknown, path exploration virtually considers $r$ as a symbolic input and 
has to explore each of the possible
dispatch targets of the call, while only one target would be explored
if the type information were provided. This causes many spurious paths to be explored,
and even renders the symbolic execution of programs that contain many
virtual function calls intractable. 

Previous work seeks to resolve this problem by 
running concrete execution first and 
then switching to symbolic execution when, e.g., some function of interest is invoked~\cite{combined-spf}.
It utilizes the execution context generated by concrete execution for symbolic execution, which is a promising
direction. Nevertheless, the path exploration is severely limited in the previous work,
because all variables inside the context are regarded as concrete inputs (even though some of
them should be symbolic iputs); we call it \emph{over-constrained symbolic execution}.
How to properly leverage concrete execution for symbolic execution is still an open research problem. 


Third, to implement the symbolic executor, a straightforward design is to place it 
inside the Android system. But this way the symbolic executor 
is tightly coupled with Android. The implementation 
has to handle compatibility with the components of the Android system,
such as the Android Runtime (ART). This significantly complicates the implementation.
Moreover, whenever the Android system is updated,
the implementation has to be maintained. To avoid the complicated implementation
and endless maintenance, a decoupled architecture is needed.

Finally, in order to implement an exploit generation
system, a large number of engineering issues
have to be addressed. For example, it has to handle class loading and system bootstrapping, 
deal with the inter-service communication and calls to native code, 
and construct exploits using path conditions. The process
of addressing these engineering issues involves much innovation as well as tremendous effort.

\noindent \textbf{\emph{Approaches} \enspace} 
Instead of tracking how variables are derived from a malicious app,
we seek to identify them by monitoring how they are used.
Based on the insight
into the patterns how variables in Android Framework
are accessed, we propose \emph{slim} tainting to capture the access
to variables derived from the malicious app and identify them as symbolic inputs precisely.
Compared to conventional tainting that usually requires significant
manual effort to handle overtainting and undertainting, the slim
tainting is automatic and precise (Section~\ref{sec:symbolic}).

We run the initialization phase of Android Framework as whole-program concrete execution,
and then perform symbolic execution
 under the context provided by the initialization phase.
This is based on the observation of Android Framework that it consists of
the initialization phase and then the ready-for-use phase, and 
the initialization phase is fairly stable among different runs as long as the system
configuration does not vary. 
The combination of concrete and symbolic executions avoids the state space
explosion due to the complex initialization and meanwhile does not lose
the execution context (Section~\ref{sec:concrete}). 
To avoid over-constrained symbolic execution,
in the execution context provided by the concrete execution, variables 
derived from the malicious app are identified by weaving slim tainting
into symbolic execution, such that these variables are specified as symbolic inputs
just in time during symbolic execution.


An architecture that puts the symbolic executor into the 
Android system leads to a coupled implementation. In contrast, 
the symbolic executor in our design is placed
out of the Android system, implemented in a way independent from
the Android system components.
The novel architecture allows concrete execution to be run on Android
and symbolic execution on the independent symbolic executor.
Now, the problem is reduced to how the symbolic executor
recognizes the information in the execution context provided by 
concrete execution (Section~\ref{sec:decoupled}).


We have overcome the scientific and engineering challenges, and implemented the system named \aeg.
We utilize \aeg to construct 
new bug finding techniques.
Given a security property $P_s$, \aeg automatically finds possible
paths in Android Framework where $P_s$ may be violated: 
the violation of $P_s$ is represented as a constraint added to each path condition, 
and if the augmented path condition is resolvable, a possible vulnerability is found.
The new bug finding techniques are automatic and
guarantee zero false positives, in contrast with recent research on
finding Android Framework bugs that requires laborious and error-prone manual work~\cite{kratos, task}.
Besides, \aeg generates hundreds of exploits in minutes, and the exploits are verified 
on different versions of Android systems.
We report our new findings on bugs, exploiting conditions, and 
more accurate vulnerability description.

\begin{itemize}
\item 
\aeg is the first system that performs automatic generation of
exploits that take advantage of Android Framework vulnerabilities,
and the first system that
supports symbolic execution of Android Framework.
It significantly changes the
state of the art in finding Android Framework bugs and exploiting them.

\item Unique challenges that cannot be resolved by existing
symbolic execution techniques have been identified; specifically,
how to identify symbolic inputs given a complex information flow from an app,
how to leverage concrete execution without leading to over-constrained
symbolic execution, and how to design a symbolic executor decoupled from the target system. 

\item We present novel ideas and techniques to address the challenges
in the context of Android Framework, such as a new approach
to combining concrete execution and symbolic execution,
and an architecture that decouples the symbolic executor from the target system.

\item We have implemented \aeg after overcoming many research 
and engineering challenges, and evaluated it in terms of effectiveness and precision.





\end{itemize}

\section{Background and Overview} \label{sec:overview}

\subsection{Background}

Android Framework provides a collection of system services, which
implements the fundamental features within Android, such as managing
the life cycle of all apps, organizing activities into tasks, and managing
app packages.  Most of the system services, except for the media services, 
run as threads in the System Server process~\cite{framework}. Thus, the System 
Server process plays a central role in Android Framework.  This work uses the 
services in this process as examples to illustrate the ideas and techniques,
which should be applicable to other services.


A service exposes \emph{service interfaces}, which are APIs invokable from apps, 
by declaring them in an Android Interface Definition 
Language (AIDL) file~\cite{aidl}. When an app invokes a service API, the call is 
passed through the IPC mechanism Binder and handled 
by the process hosting the service. 

\subsection{Overview}
Given a specific security property $P_s$, a malicious app invokes one of the service interfaces
to drive the Android Framework execution to violate $P_s$.
Therefore, exploit generation leads to three questions: Which service interfaces should be invoked?
What conditions in terms of the parameter values and the app configuration
should be satisfied, such that
the invocation leads to a successful exploitation? How to build a malicious app based on the conditions?
The questions are resolved by the following three steps, respectively.


The service interface method that should be invoked to launch an attack is
called the attack's \emph{entrypoint service interface}.
Given $P_s$, it is usually straightforward to determine
the entrypoint service interface for launching the attack. 
For example, if the violation of $P_s$ refers to that
a service interface method accesses some resource without sufficient permissions, then this service interface
is the entrypoint one. 
In addition to making using of API specification and expert knowledge,
static analysis can also be used to determine the entrypoint. For example,
assume that a DoS attack is launched by raising unhandled exceptions from some internal 
method, then service interfaces that can reach the internal method are
entrypoints. Thus, the selection of the entrypoint mainly depends on
the attack, and we will discuss this step when concrete attacks are 
considered in the evaluation (Section~\ref{sec:evaluation}). We do not bind our 
system to any specific selection methods.
The second step is to obtain the app configuration values
and the parameter values used to invoke the entrypoint service interface. 
Our approach is to utilize symbolic execution to find such input values, 
which is the focus of this paper.
Finally, the values obtained at the second step are used
to build exploits. Our evaluation demonstrates 
how to build exploits from the values. 





\begin{figure*}[!htb]
\centering
\aaf 
\includegraphics[scale=0.75]{overview.pdf}
\captionsetup{font={small}}
\aaf \aaf  \af
\caption{Architecture of \aeg.} 
\aaf \af
\label{fig:arch}
\end{figure*}

Figure~\ref{fig:arch} shows the architecture of \aeg. At its core is the
symbolic execution engine, which performs symbolic execution of Android Framework.
The output is used to build exploits.
Between the engine and the Android system is the execution
context query server, which migrates the execution context information from 
Android to the engine. The details 
will be described in the following sections.

\subsection{Skeleton Malicious App} \label{sec:skeleton}
There is a dilemma that we are in the process of completing 
a malicious app, while the malicious app has to get launched
so that we can identify variables in Android Framework derived from the app. 
But how to launch a malicious app that has not been completed? 
Our insight is that, due to the nature of symbolic execution,
the concrete values of symbolic inputs do not matter (detailed in Section~\ref{sec:symbolic}); 
hence, a \emph{skeleton malicious app} (skeleton app, for short) that
works as a placeholder suffices. 
The skeleton app can be an app selected 
to carry the exploit. 
We assume that the skeleton app is specified or provided by users of our system (i.e., malware
developers). Or, for security analysts' purpose,
it can be one that contains
all the aspects of a regular app, including the manifest file,
activities, and services, but does not implement any essential functionality; 
in particular, the skeleton app used in our experiments borrows the
manifest file from the Android developer 
website, which has ``every element that it can contain''~\cite{manifest}.

\section{Combining Concrete \& Symbolic Executions} \label{sec:concrete}

\subsection{Missing Execution Context} \label{sec:ucse}


Android Framework has a complex initialization phase before system
services are ready for use. 
However, symbolic execution suffers from the path explosion problem, as 
the number of distinct execution paths is often exponential in the number of branches. 
Thus, symbolic execution starting from the program entry of Android Framework \texttt{SystemServer.main}  
probably fails to finish the initialization phase. 
Besides, the initialization creates multiple threads for running services,
which also complicates the symbolic execution.
Therefore, symbolic execution that can skip the complex initialization phase
is desirable.
 


\begin{definition}
An \emph{execution context} consists of the program counter, register file, stack, and heap.
For a Java program, the heap is a collection of classes and objects. 
\end{definition}

Under-constrained symbolic execution can directly start from
an arbitrary function within a program~\cite{ucklee}. 
As it effectively skips the costly initialization phase, this approach reduces the 
number and length of execution paths to be explored. 
However, due to skipping the functions in the path prefix,
the execution context for symbolic execution is missing. Specifically,
the type and value information of the \emph{input variables}, 
i.e., non-locally defined variables read during symbolic execution,
is lost. It causes several problems.



First, without the type information, it is hard to determine the dispatch target
of a virtual function call.  
Consider an example \texttt{s.iterator()}, where \texttt{s} is a reference of
the \texttt{Set} interface type. 
But \texttt{Set} is implemented by over 40 subclasses in Android Framework code, which means that
symbolic execution needs to try each possibility, causing many spurious paths to be explored.
Note that such virtual function calls prevail in Android Framework.


Second, due to the lack of the value information of variables, it is unknown how to handle
instructions or function calls that involve them. For example,  
consider \texttt{LocationManagerService.mProviders}, which is an ArrayList
that stores currently installed GPS providers;
as the elements and the length of the ArrayList are unknown, it is
hard to carry out a loop that iterates through the list. One workaround
is to regard the list as a symbolic input and then handle it using lazy initialization~\cite{lazy};
this way, however, the loop becomes unbounded and elements of the list
become symbolic, which exacerbates the path explosion problem.

\subsection{Solution} 


Our observation of Android Framework is that its execution consists
of the initialization phase and the ready-for-use phase, and the
initialization phase is fairly stable among different runs, since
the system boots according to the system configuration and the currently
installed apps, which are stable.
We thus \emph{run the initialization phase as whole-program concrete execution,
and then perform symbolic execution starting at the entrypoint service interface method}
under the execution context provided by the concrete execution. 
It is notable that the type and value information of
all the variables is available, which directly resolves
the issues discussed in Section~\ref{sec:ucse}.
The combination of concrete and symbolic executions avoids the state space
explosion due to the complex initialization, and meanwhile
preserves the execution context for symbolic execution. 

For the purpose of exploit generation, the initialization phase
refers to both the system initialization and the initialization
of the skeleton malicious app until it is ready to make calls to
the entrypoint service interface method, such that Android 
Framework has all the variables derived from the skeleton app
that can possibly be used by the service interface call.

However, the concrete execution leaves an execution context
where every variable is concrete, including those derived
from the skeleton app. How to perform symbolic execution 
in this situation? Previous work that switches from
concrete execution to symbolic execution simply uses the
concrete values of the variables in the execution context and
only considers the parameters of the function-under-test as symbolic inputs~\cite{combined-spf}, 
which severely limits path exploration and leads to over-constrained
symbolic execution. Unlike the previous work, we identify variables derived from the skeleton app 
as symbolic inputs,  such that the path exploration considers 
all possible values of the variables, as presented
in the next section.


\section{Identifying Symbolic Inputs} \label{sec:symbolic}
When symbolic execution is applied to exploit generation targeting stand-alone
executables, the form of exploits is usually simple,  
e.g., the SQL query string used to launch SQL injection attacks and 
the http request used to exploit a buffer overflow bug in a web service.  
Variables corresponding to these exploit inputs are usually easy to identify.
In our case, however, given the complex information
flow from the skeleton app to Android Framework, it is difficult to identify which variables 
in the execution context have been derived from 
the skeleton app (Section~\ref{sec:scatter}).
We describe some straightforward thoughts on solving the problem (Section~\ref{sec:attempts}), 
and then present an automated and precise solution (Section~\ref{sec:symbol-solution}). 

\subsection{Scattered Symbolic Inputs} \label{sec:scatter}

Android Framework provides a running environment for all apps, each of which has
information stored in Android Framework, such as granted permissions, 
activities, and intents. There are also variables allocated for system services. 
The variables used to store the information of 
different apps and system services are mixed together in the memory address space, and
there is no clear boundary between variables derived from the
skeleton app and other variables. 

A closer look at the Android Framework design shows that
there are two distinct types of variables. The first
type, called \emph{non-app-specific} variables,
includes those that are allocated and maintained regardless of specific apps in the system.
For example, the aforementioned \texttt{LocationManagerService.mProviders} ArrayList
 is maintained regardless of specific apps.
As another example,
\texttt{AudioService.mConnectedDevices} is a HashMap
that stores currently connected devices and does not depend on specific apps either.  
There are many other variables belonging to this type; for instance, a set of field variables 
(\texttt{mState}, \texttt{mNetworkType}, \texttt{mTypeName}, etc.) 
in the \texttt{NetworkInfo} class that describe the statuses of a network interface.
\emph{Non-app-specific variables should be used as concrete inputs
rather than symbolic ones, as a malicious app typically does not have the capability
of manipulating such system data.}

The second type, called \emph{app-specific} variables, 
stores app-specific information.  
Some variables store information  for all the installed apps;
for instance, \texttt{Settings.mUserIds} 
is an ArrayList that stores the installation data of each installed app 
(the code path, signature, first install time, last update time, granted permissions, etc.). 
Others store the information of running apps, such as task affinities, 
intents, and back stacks.

Unlike the Linux kernel, which stores most information of
a process in a centralized structure \texttt{task\_struct},
it is notable that the app-specific information is organized
according to the system services (rather than apps) probably 
because Android Framework is programmed as a set of system service classes.
Given an app, its related information scatters in many different collection
data structures in Android Framework. 

The task of selecting symbolic inputs is to
find the app-specific variables and to locate elements 
derived from the skeleton app within the variables. E.g., 
in addition to determining \texttt{Settings.mUserIds} is 
an app-specific variable, we need to locate which element in the array is derived from
the skeleton app, like looking for a needle in a large pile of hay.


\subsection{Thoughts on Tainting} \label{sec:attempts} 
Tainting is a natural approach that may be used to track
the information flow from the skeleton app. By specifying all the return values of
function calls that read the
apk file of the skeleton app as the taint source,
all the tainted variables in the execution context must have
been derived from the skeleton app. 
However, such information
flow involves multiple intricate steps, including app installation,
system boot, and starting the app. Specifically, after an app is 
installed, its code and data are stored into multiple files in the system, which 
are parsed and read at different stages during booting the system 
and starting the app. 


Given the complexity of these steps, it is very unlikely to precisely tracking
the information flow throughout these complex steps.
First, without taint sanitization, more and more values would become
tainted, which results in overtainting~\cite{all-taint}. 
How to insert sanitization properly has been a challenging problem,
especially considering the codebase and complex logic of the Android system. 
In addition, undertainting can arise, for example, when
information flow occurs through control dependencies, 
tainting based on data flow only is inadequate and needs dedicated handling. 
Therefore, a comprehensive and precise tracking of the information from the skeleton
app is hard to achieve. 



\subsection{Slim Tainting for Identifying Symbolic Inputs} \label{sec:symbol-solution}

\subsubsection{Locating the Needles}

Our investigation of Android Framework reveals that 
app-specific variables are stored in two categories of data structures:
array-based ones (built-in arrays, ArrayList, SparseArray, etc.) and
 hash-table-based ones (HashMap, HashSet, etc.). Given an app, 
the app's corresponding element from an app-specific data structure 
is retrieved in one of the two characteristic ways.

\begin{figure}[t]
\caption{Example of retrieving information from an
array-based variable.} \label{fig:two}
\vspace*{6pt}
\begin{lstlisting}[language=Java,escapechar=|]
// static final int PER_USER_RANGE = 100000;
// static final int FIRST_APPLICATION_UID = 10000;

// Defined in the PackageManagerService class
// The caller first sets "uid = Binder.getCallingUid();"
int checkUidPermission(String permName, int uid){
  r = mSettings.getUserIdLPr(UserHandle.getAppId(uid));
  ...
}
// Defined in the UserHandle class
static final int getAppId(int uid) {
  return uid % PER_USER_RANGE;
}
// Defined in the Settings class
ArrayList<Object> mUserIds;
Object getUserIdLPr(int uid) {
  if (uid >= Process.FIRST_APPLICATION_UID) {
    int index = uid - Process.FIRST_APPLICATION_UID;|\label{ln:index}|
    ...
    //index = uid%100,000-10,000
    return mUserIds.get(index); 
  } 
}
\end{lstlisting}
\aaf \af
\end{figure}

\begin{figure}[t]
\caption{Example of retrieving information from a
hash-table-based variable.} \label{fig:three}
\vspace*{6pt}
\begin{lstlisting}[language=Java,escapechar=|]
// Defined in the PackageManagerService class 
HashMap<String, PackageParser.Package> mPackages;
int checkPermission(String perm, String pkgName){
  PackageParser.Package p = mPackages.get(pkgName);
  ...
}
\end{lstlisting}
\aaf \af
\end{figure}

First, given an array-based variable, the Android Framework program 
retrieves an app's information in the array 
using an index that is a function of the app's UID (an app's UID is assigned upon installation and not changed).
Our investigation shows that there are only two such formulas used to calculate the index. 
One is $(uid\%100,000-10,000)$, converting the 
user app's UID into an index to retrieve the element for the app 
from a built-in array or ArrayList;
the other one is $(uid\%100,000)$, which is used to calculate 
the index into a SparseArray.
For example, as shown in Figure~\ref{fig:two}, the first
formula is utilized to calculate the index 
into the ArrayList \texttt{Settings.mUserIds}, which stores the information of all the
installed apps with one element for each app.

Second, for hash-table-based variables, the package name
(or the package name concatenated with a component name) is used as the key
to access elements. Figure~\ref{fig:three} shows such an example. Since the skeleton app has 
a unique UID and package name, they are used as taint sources to track the
access to variables derived from the skeleton app. 


Given the execution context provided by the concrete execution,
we seek to identify variables derived from the skeleton app,
and specify them as symbolic inputs.  
To achieve it, \emph{slim tainting} is proposed to recognize 
the characteristic access patterns discussed above, and the tainting is weaved
into symbolic execution to specify the identified variables as symbolic inputs just in time.
Specifically, the return value of \texttt{getCallingUID} and the 
package name of the skeleton app are set as taint sources.
The taints propagate through (1) the modular and subtraction 
operations on the UID, and (2) the string
assignment and concatenation operations involving the package name.
The taint sinks include the \texttt{get} functions of the collection data structures
 as well as bytecode instructions for loading elements from built-in arrays, such as \texttt{iaload} and \texttt{aaload},
which check whether the index or key being used has been tainted; if so, the corresponding
element is set as a symbolic input.

\begin{figure}[ht]
\centering
\af
\hspace*{-9pt}
\subfloat[ArrayList \texttt{mUserIds} using $(uid\%100,000-10,000)$ as indexes.]{%
	\includegraphics[scale=0.6]{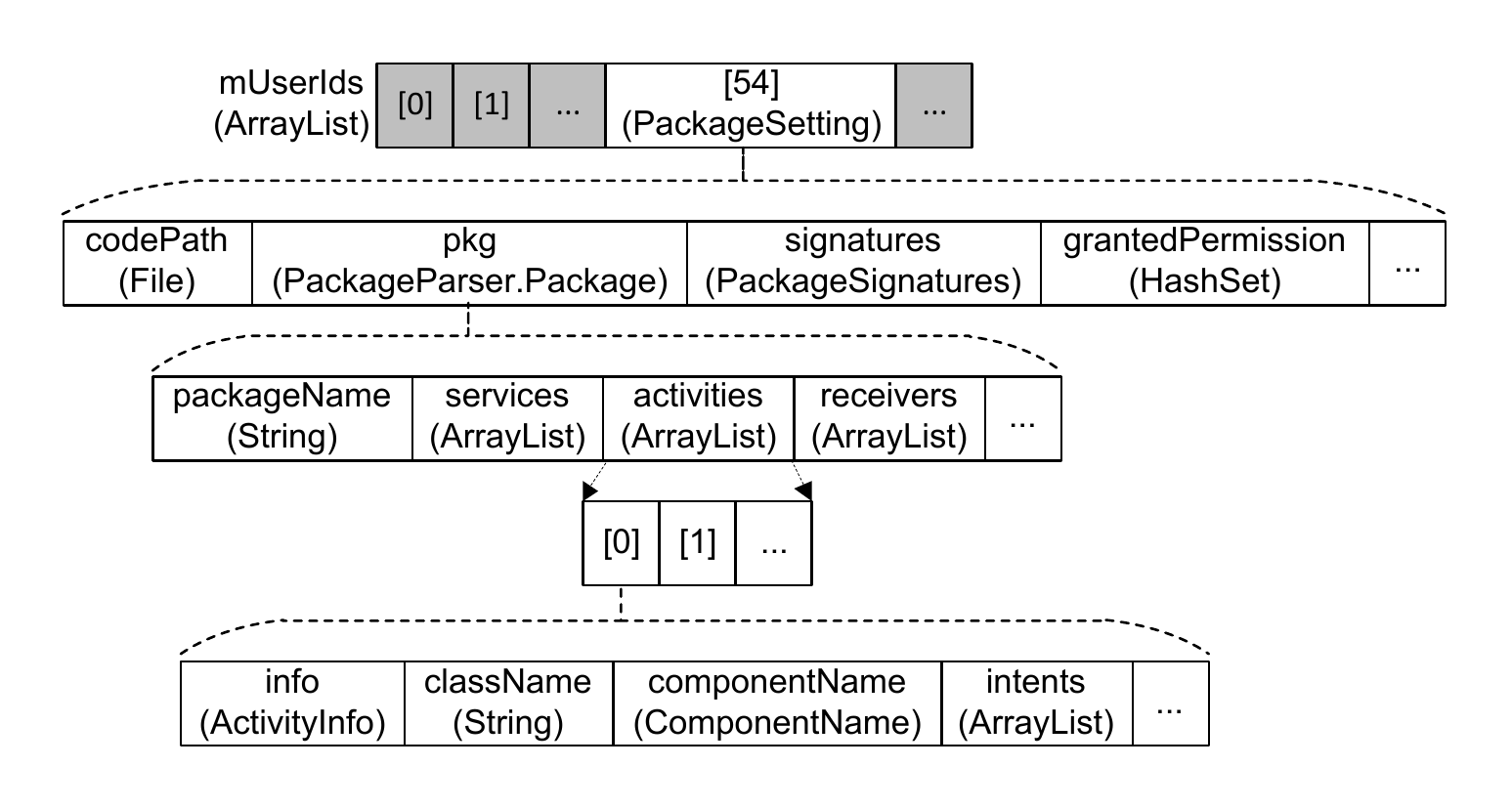}
	\label{fig:array}%
	}
\aaf

%
\hspace*{-6pt}
\subfloat[HashMap \texttt{mPackages} using package names as keys.]{%
	\includegraphics[scale=0.6]{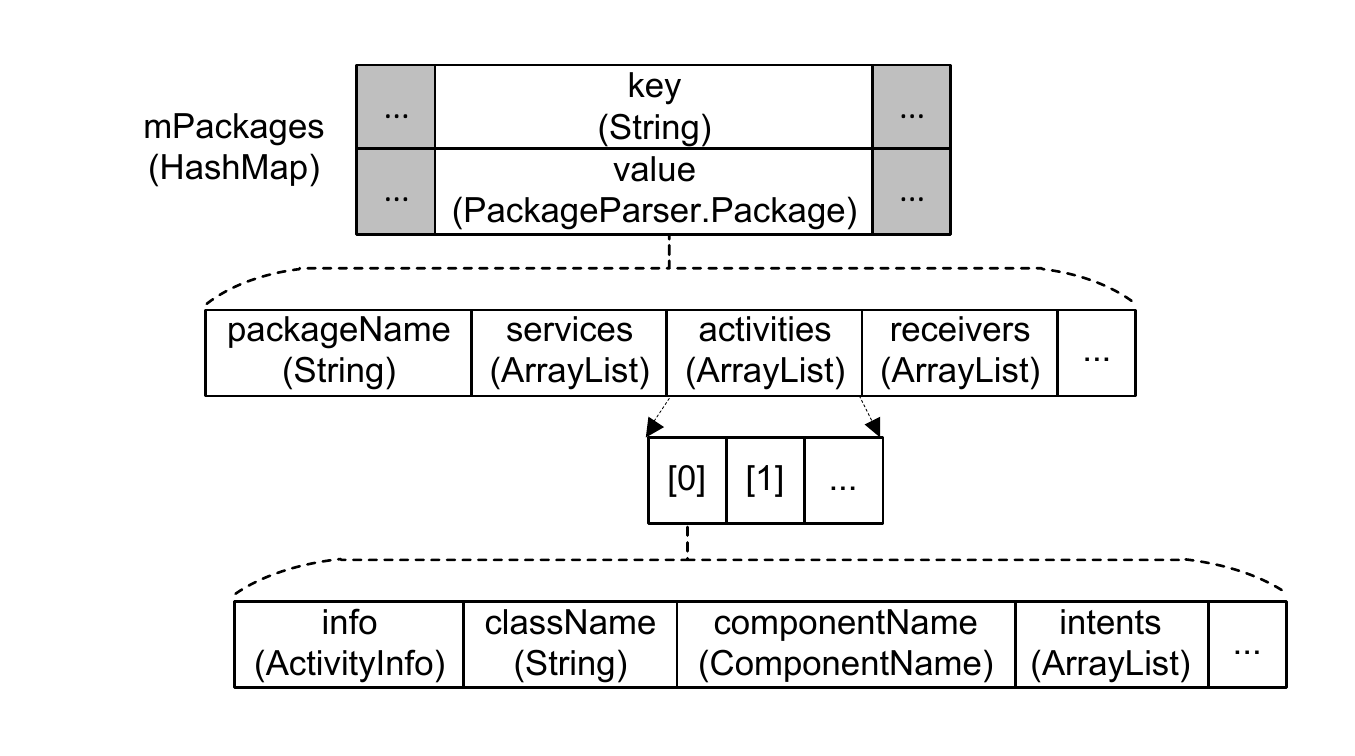}%
	 \label{fig:hashtable}%
	}
\caption{Examples of app-specific variables.} \label{fig:zzz}
\aaf 
\end{figure}

Figure~\ref{fig:zzz} shows two examples. Figure~\ref{fig:array} 
depicts an ArrayList variable \texttt{mUserIds}. 
The names within the parentheses are the types of the corresponding 
variables. Assume the skeleton app UID is 10,054. Due to dynamic taint propagation,
the calculated index 54 (= $(uid\%100,000-10,000)$) is tainted; hence, the corresponding 
element in \texttt{mUserIds} is set as a symbolic input when it is loaded using \texttt{aaload}. 
Figure~\ref{fig:hashtable} shows a HashMap variable
\texttt{mPackages}, which uses package names as keys. 
Since the package name of the malicious app is a taint source,
when it is used to retrieve an element from \texttt{mPackages},
the element is identified as a symbolic input when it is retrieved using the HashMap's \texttt{get} function.


Slim tainting involves very specific taint sources and operations,
avoiding the overtainting and undertainting problems in
conventional tainting, and requires no manual effort.
 


\subsubsection{Relations between Symbolic Inputs}

Android Framework contains a large number of complex, pointer-rich data structures. 
Figure~\ref{fig:zzz} shows such examples. Each element in the ArrayList (Figure~\ref{fig:array})
is a reference to a \texttt{PackageSetting} instance, which comprises
many references  to other complex variables, e.g., a  
\texttt{PackageParser.Package} class instance; it is notable that
symbolic inputs in both the ArrayList (Figure~\ref{fig:array}) and the
HashMap (Figure~\ref{fig:hashtable}) point to 
the same \texttt{PackageParser.Package} instance directly or indirectly.

Such relations are hard coded in Android Framework, and
it is critical to take the relations into account during symbolic execution. For example,
if the two symbolic input references in Figure~\ref{fig:zzz} were regarded as independent, then
each would be considered to be pointing to a separate \texttt{PackageParser.Package} instance;
constraints in the path condition that should have described the single \texttt{PackageParser.Package} instance
would be split to describe two instances, which is incorrect.

Existing symbolic execution techniques have not resolved
the problem of describing the relations between symbolic inputs automatically. 
Lazy initialization can be used to synthesize a data structure,
but still relies on manual effort to write specification about the data structure~\cite{lazy}.

To resolve the problem, the concept of \emph{semi-symbolic reference} is proposed for handling 
reference-typed symbolic inputs. 
Specifically, let $r$ be a reference that is identified as a 
symbolic input through tainting, and $o$ be the object pointed to by $r$.
When $r$ is used to access $o$, the symbolic executor first
examines whether \texttt{o.symbolicHandled} is true, where \texttt{symbolicHandled} is a flag
added to each object indicating whether the host object has been identified as a symbolic input.
If it is \texttt{false}, all the primitive-typed fields in $o$ 
are set to conventional symbolic inputs, while other reference-typed fields are set to
semi-symbolic references (for recursive handling), and \texttt{symbolicHandled}
is set to \texttt{true}; otherwise, no handling is performed to avoid duplicate processing. 

There are several critical points in the solution. First, 
the solution benefits from the execution context information:
when two references point to the same object, they have the same
reference value; thus, when an object is processed multiple times due to multiple
references pointing to it, the solution can recognize that they
point to the same object and ensure the object is identified as a symbolic input only
once (based on the \texttt{symbolicHandled} flag).
For example, in Figure~\ref{fig:zzz}, after the \texttt{PackageParser.Package} object
is handled once, its \texttt{symbolicHandled} flag must be \texttt{true}.
Second, a semi-symbolic reference propagates this attribute
to all the reference-typed fields in the object it points to, such that they are
handled recursively. The propagation is valid as in Android Framework each element in an app-specific
data structure is a ``cell'' that stores information for a specific app;
this design ensures that once the execution obtains the reference to some 
element in an app-specific data structure, the subsequent
access is bound to the information of the app stored in that element without worries that the access
may reach another app's information.

\section{Decoupled Architecture} \label{sec:decoupled}
By combining concrete execution and
symbolic execution,
the state space explosion due to the complex initialization is avoided, but 
the hybrid execution idea requires a more careful design. This section 
first describes the design choices we made, 
and then presents our design.

\subsection{Design Choices}
A straightforward design is to augment the
Android system to add the capability of symbolic execution by modifying Android Runtime (ART).
It takes advantage of the ART's capability of concrete execution, but
requires a lot of modifications to enable symbolic execution, which
is very different from concrete execution in terms of thread management, garbage collection,
object representation, instruction execution, etc. Therefore, it is challenging to make the
two types of execution coexist in the same system. Not only does the compatibility
with concrete execution have to be handled carefully, but it implies endless effort to maintain the
implementation for the frequently updated Android system.




Instead of implementing a coupled system,
we propose to allocate the two kinds of executions to two systems: an
original Android system for concrete execution and the other system outside Android for
symbolic execution. As the latter is specialized for symbolic execution, 
its design and implementation are largely simplified.  
Moreover, since the symbolic execution engine is decoupled from the
Android system, it does not need to be maintained
when the Android system is updated.


\subsection{Migration Algorithm} \label{sec:alg}

Symbolic execution is launched by executing
a driver program that invokes an entrypoint system interface method. Upon starting, 
the program counter, register file, and stack all obtain their fresh
content, while the heap, which is a collection of classes and objects, 
needs to be migrated from the execution context 
provided by the concrete execution.


\begin{algorithm}[!tb]
\small
\captionsetup{font={small}}
\caption{Migration of heap information.}

\label{alg:getfield}
\begin{algorithmic}[1]

\Function{getfield}{index}
   \State objRef = peekStackTop() \label{ln:peek}
   \State fdInfo = getFdInfo(index) \Comment{Class-specific info.} \label{ln:fdinfo}
   \State fd = getFd(objRef, fdInfo) \Comment{\emph{objRef}-specific info.} \label{ln:fd}
   \If{!fd.getSnapshotRefAttribute()} \label{ln:ismapped}
      \State return super.getfield(index)
   \EndIf
   \State concRef = fd.getValue() \label{ln:value}
   \State symRef = conc2Sym.get(concRef) \label{ln:conc2sym}
   \If{symRef == NULL} \label{ln:symnull}
      \State fdType = fdInfo.getFdType() \label{ln:fdtype}
      \If{fdType == strRef} \label{ln:str1}
         \State str = snapshot.getStr(concRef)
         \State symRef = searchConstantPool(str);
         \If{symRef == NULL}
            \State symRef = newString(str);
         \EndIf \label{ln:str2}
      \ElsIf{fdType == arrayRef} \label{ln:array1}
         \State entryType = fdType.getEntryType()
         \State len = snapshot.getArrayLen(concRef)
         \State symRef = newArray(entryType, len)
         \State snapshot.copyEntries(symRef, concRef) \label{ln:array2}
      \Else \Comment{Other reference types} \label{ln:obj1}
         \State symRef = newObj(fdType) \label{ln:newobj}
         \State snapshot.copyFields(symRef, concRef) \label{ln:obj2}
      \EndIf  \label{ln:end}
      \State conc2Sym.addPair(concRef, symRef) \label{ln:addmapping}
   \EndIf
   \State fd.setValue(symRef) \label{ln:setvalue}
   \State fd.setSnapshotRefAttribute(false) \label{ln:setfdmapped}
   \State return super.getfield(index)
\EndFunction
\State
\Function{initClass}{classInfo}
   \If{snapshot.isInitialized(classInfo)}
      \State snapshot.copyStaticFields(classInfo) \label{ln:static}
   \Else
      \State super.initClass(classInfo)
      \State handleBootstrapField(classInfo) \label{ln:bootstrap}
   \EndIf
\EndFunction
\end{algorithmic}
\end{algorithm}

\begin{table}
\caption{Bytecode instructions (and function) used for migrating heap information.} \label{table:instruction}
\centering
{\small
\scalebox{0.94}{
\begin{tabular}{ c | c | c }
\hline
\multirow{2}{*}{Instruction} & Stack & \multirow{2}{*}{Description} \\
                             & [before]$\rightarrow$[after]  &  \\
\hhline{=|=|=}
\texttt{getfield} & objRef $\rightarrow$ value & get a field value of an object\\ 
\hline
\texttt{getstatic} & $\rightarrow$value & get a static field value of a class\\
\hline
\multirow{2}{*}{\texttt{aaload}} & arrayRef, index & load onto the stack a reference  \\ 
                                 & $\rightarrow$ value & from an array \\
\hline
\texttt{initClass} & N/A & invoked for class initialization\\ 
\hline
\end{tabular}
}} \af
\end{table}

The heap memory image in the execution context provided by the concrete execution
is called a \emph{snapshot}. We present an algorithm that migrates classes and objects
from the snapshot to the JVM for symbolic execution. 
Whenever a class or object is referenced, it is migrated from
the snapshot by allocating space in the symbolic executor and then copying the fields. 
The algorithm is built into the symbolic execution engine, which interprets 
the Java bytecode (of Android Framework) in a non-standard way. 
Algorithm~\ref{alg:getfield}  shows the main migration procedures, 
which override the interpretation of several bytecode instructions. 
Table~\ref{table:instruction} shows the list of bytecode instructions whose interpretation is overridden
to support migration; for each instruction, the effect that the 
instruction has on the operand stack and the description are included.
We first introduce a data structure and a flag that are important for 
migration, and then describe how objects and classes are migrated.

\noindent \textbf{Migration hash table.} 
A hash table, \texttt{conc2Sym}, is maintained to map
reference values in the concrete execution world (where the snapshot has been captured) 
to ones in the symbolic execution world. Every time an object $o$ is 
migrated, a new pair $\langle r_c, r_s \rangle$ is added to the hash table, where $r_c$ is 
the reference value of $o$ in the concrete execution world and $r_s$
symbolic. The hash table is maintained for two purposes. 
First, it prevents duplicate migration of an object; that is, an object pointed to by $r_c$ is migrated
only if $r_c$ is not found in the hash table. Second,  the hash table is used to
translate reference values in the concrete execution world, if they exist in the hash table, 
to ones in the symbolic execution world. The hash table is handled as part of the process state,
and gets stored and restored as the path exploration advances and backtracks, respectively.

\noindent \textbf{Reference flag.} A flag \texttt{snapshotRef} 
is associated with each reference-typed field (and each reference-typed element in 
an array, as well) by the symbolic executor to indicate whether its value
is a reference value in the concrete execution world.  When an object is newly migrated, 
the \texttt{snapshotRef} flags of all its reference-typed fields (and elements in an array object) 
are set to \texttt{true}, since the field values only make sense in the concrete execution world.
Once a field is updated with a reference value in the symbolic execution world, 
its \texttt{snapshotRef} is set to \texttt{false}. 

\noindent \textbf{Migrating objects.}
Given a reference to an object on the stack (Line~\ref{ln:peek}), 
the instruction \texttt{getfield} pushes a field value of the object onto the stack.
If the field's \texttt{snapshotRef} attribute is \texttt{false} (Line~\ref{ln:ismapped};
note that, for all primitive-typed fields, \texttt{getSnapshotRefAttribute} returns \texttt{false}),
which means that either it is a primitive-typed field or it has a reference value in the symbolic execution world,
the instruction's interpretation is not changed. If \texttt{snapshotRef} is \texttt{true} 
and the field value \texttt{concRef} is not found in \texttt{conc2Sym}  (Line~\ref{ln:symnull}), 
the object should be migrated (Lines~\ref{ln:fdtype}--\ref{ln:end}); after migration, 
the pair $\langle concRef, symRef\rangle$
is added to \texttt{conc2Sym} (Line~\ref{ln:addmapping}).

How to migrate an object is determined by the object type (Line~\ref{ln:fdtype}).
(1) If the object is a string, the algorithm first searches for a string that has the same
value within the runtime constant pool (which stores a set of string literals) in the 
VM for symbolic execution.  If not found, a new string
with the same value is created in the symbolic world (Lines~\ref{ln:str1}--\ref{ln:str2}).
(2) If the object is an array, an array is allocated and all the elements are copied to the new
array (Lines~\ref{ln:array1}--\ref{ln:array2}). This algorithm performs a shallow copy. Thus,
for a multi-dimensional array, e.g., $A[5][10]$, only the 5 elements in the top-level array are copied
at this moment. Later, to access any of the 5 elements, the instruction
\texttt{aaload} is executed, which is the reason the interpretation of \texttt{aaload}
is also overridden to migrate second-level arrays (not shown in Algorithm~\ref{alg:getfield}). 
This reflects the principle of lazy migration: an array object is not copied until a
reference to the object is accessed. 
(3) A reference to an ordinary object is handled by allocating a new object and copying all its
fields (Lines~\ref{ln:obj1}--\ref{ln:obj2}).


While non-static fields are accessed through \texttt{getfield}, access to static fields
is through \texttt{getstatic}. Thus, to migrate objects pointed to by static fields, 
the interpretation of \texttt{getstatic} has to be overridden, and the interpretation 
is similar to that of \texttt{getfield}. 

\noindent \textbf{Migrating classes.}
When an operation (e.g., an object of a class is created or a class's static fields are accessed for the first time) 
triggers initialization of a class during symbolic execution, 
\texttt{initClass} is invoked by the underlying VM for symbolic execution automatically.
For classes that have been initialized during concrete execution, the symbolic
executor has to make sure that they are migrated instead of being initialized, considering that the static fields
have obtained their values during concrete execution. 
Thus, when \texttt{initClass} is invoked,
the symbolic executor first checks whether the class has been initialized in the
concrete execution world; if so, the enclosed static
fields in the class are copied from the snapshot to the symbolic execution 
world (Line~\ref{ln:static}). In particular, when an object of some class
is created in the symbolic world for the first time
due to migration (Line~\ref{ln:newobj}), it triggers the invocation of \texttt{initClass} first, 
which migrates the class.

\subsection{Bootstrapping}   

An important invariant kept during migration
is that, whenever a field of an object $o$ (resp.\ an element of an array $A$) is accessed, 
$o$ (resp.\ $A$) must have existed in the symbolic execution world. The invariant is achieved because, 
upon accessing a reference-typed field, the object pointed to by the field gets
migrated. Assume $f$ is the field whose access triggers the migration
of the first object; a natural question is where $f$ resides. This question
is resolved in the test driver. 

\begin{figure}
\caption{Example of a test driver.} \label{fig:boot}
\begin{lstlisting}
public TestDriver() {
   @fromSnapshot
   private static com.android.server.LocationManagerService mService;
   public static void main() {
      // The parameters are configured as symbolic inputs, so their values do not matter
      mService.getProviders(null, false);
   }
}
\end{lstlisting}
\aaf\aaf\af
\end{figure}


A test driver simply invokes the entrypoint system interface method and specifies 
the \emph{bootstrap} field, whose type is a reference to the system service 
class that contains the entrypoint method. Figure~\ref{fig:boot} shows an example of
a test driver. A custom annotation \texttt{fromSnapshot} 
is used to specify the bootstrap field, which is recognized and handled by 
\texttt{handleBootstrapField} (Line~\ref{ln:bootstrap} in Algorithm 1), when the \texttt{TestDriver} class is initialized; specifically, \texttt{handleBootstrapField} 
sets the bootstrap field value to the reference value of the system service 
object in the concrete execution world, and sets the field's \texttt{snapshotRef}
attribute to \texttt{true}. (Note that all the system service classes adopt the \emph{singleton} 
design pattern, so there is no ambiguity when specifying the reference value.)
In Figure~\ref{fig:boot}, for example, when \texttt{TestDriver} is initialized, 
\texttt{handleBootstrapField} sets the bootstrap field to the reference to the 
Location Manager service object in the snapshot. Next, when the bootstrap field 
is accessed, the service object is migrated.

\subsection{Migration Tree}

\begin{figure*}[!htb]
\centering
\aaf 
\hspace*{-8pt}
\includegraphics[scale=0.54]{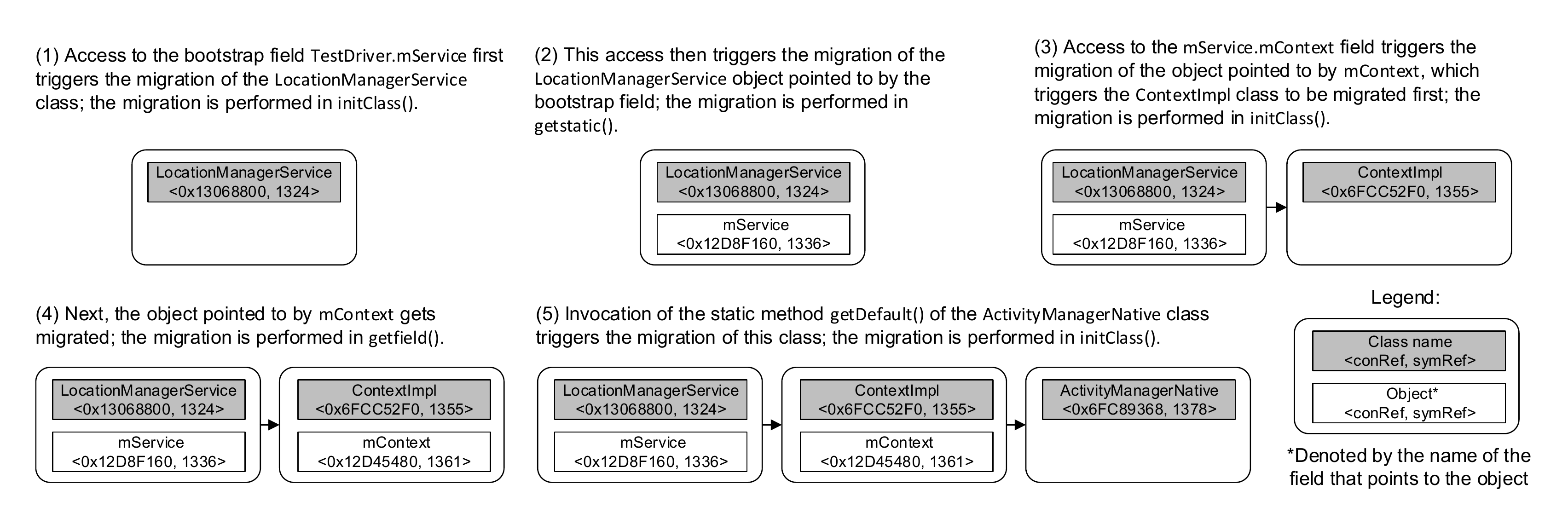}
\captionsetup{font={small}}
\aaf \aaf  \af
\caption{Process for building a migration tree. Grey rectangles
and white ones denote classes and objects, respectively. For each class
and object, \texttt{<conRef, symRef>} denotes the mapping between the reference value
in the concrete execution world and that in the symbolic world, added
to the \texttt{conc2Sym} hash table. 
} 
\label{fig:treepart}
\end{figure*}

The migration of classes and objects forms a \emph{migration tree}, which grows 
as new classes and objects are migrated, rooted at the class and object 
corresponding to the bootstrap field type.  
We use the test driver in Figure~\ref{fig:boot} as an example to illustrate how 
the migration tree is built, as shown in Figure~\ref{fig:treepart}, where
the root node is the class and object for \texttt{LocationManagerService}.
The migration of a class also triggers the migration of all its super classes, which is not shown in 
Figure~\ref{fig:treepart} for simplicity.

\begin{figure*}[!htb]
\centering
\hspace*{-20pt}
\includegraphics[scale=0.5]{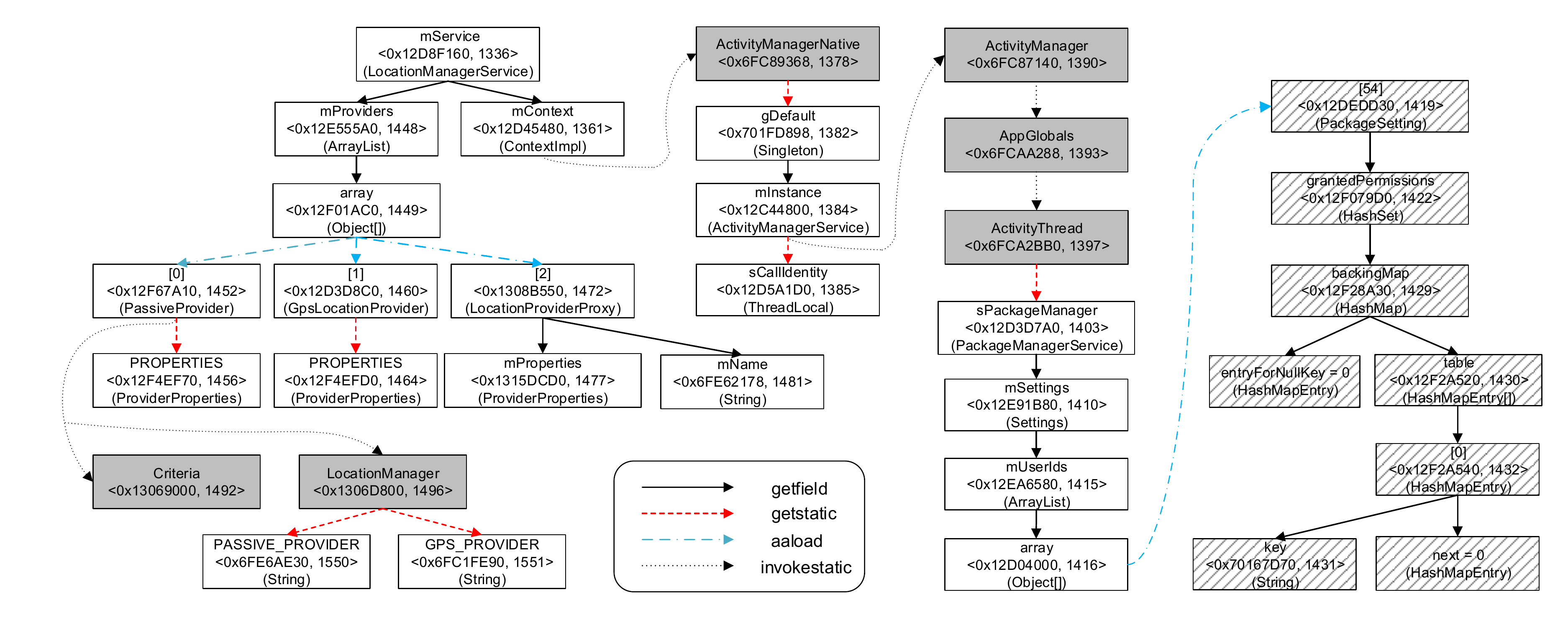}
\captionsetup{font={small}}
\aaf \aaf 
\caption{Part of a migration tree with some classes omitted. 
Different arrows are used to denote different instructions that trigger the migration.
Rectangles with diagonal stripes denote objects that are identified as symbolic inputs.} 
\aaf 
\label{fig:treeall}
\end{figure*}

Part of the resulted migration tree is showed in Figure~\ref{fig:treeall}. 
It also shows the identified symbolic inputs and how the symbolic input 
attribute propagates.
After the  element with index 54 in the \texttt{mUserIds} is identified
as a symbolic input through tainting, the symbolic input attribute is 
propagated to other variables pointed to by the element.

\section{Implementation Details} \label{sec:implementation}

We built the symbolic executor on 
Symbolic PathFinder (SPF)~\cite{spf}, 
a symbolic execution framework on top of the Java PathFinder (JPF)~\cite{jpf}.
It runs outside the Android system, and
does not rely on the Android internals, achieving the goal
of a decoupled architecture.   
This section covers important implementation details for building and configuring the system. 

\subsection{Configuration}
In addition to specifying the entrypoint system interface and the test driver, 
we need to provide the Android Framework code and the heap memory snapshot
for symbolic execution.  


\subsubsection{Classpath}
The Java source code in Android is compiled into \texttt{.jar} files, which comprise standard 
\texttt{.class} files, and the symbolic executor is built to analyze Java bytecode in such \texttt{.class} files.
The classpath below shows the classes analyzed by the symbolic executor.

\begin{lstlisting}[language=bash,numbers=none]
classpath=test_driver_dir;\
services_intermediates/classes-full-debug.jar;\
framework_intermeidates/classes-full-debug.jar;\
core-libart_intermediates/classes-full-debug.jar
\end{lstlisting}

The first line specifies the directory containing the test driver, the next two lines
the Android Framework code, and the last line the core libraries of ART, such as utility,
io, and math libraries. Several classes (e.g., \texttt{java.lang.class}, \texttt{.Thread}, 
\texttt{.StackTraceElement}) are modeled by the symbolic executor, but \texttt{core-libart}
contains the Android version of these classes; they are hence excluded 
from \texttt{core-libart} to avoid system initialization failures.

\subsubsection{Heap Memory Snapshot} \label{sec:snapshot}




After a heap memory snapshot of the System Server process is captured 
(using the \texttt{dumpheap} utility), it is first converted to a standard 
\texttt{.hprof} file using the \texttt{hprof-conv} utility included in the Android 
SDK. The standard \texttt{.hprof} file format opens up the possibility
of parsing the snapshot using many existing tools and utilities.
In our case, a HPROF heap dump parser is used to extract the list of
classes and objects stored in the \texttt{.hprof} file~\cite{hprof}.
Based on the parser, we have built an execution context query server 
that returns classes and objects requested by the symbolic executor.



\subsection{Handling Special calls} \label{sec:feature}

\subsubsection{Handling Service Calls}

\begin{figure}
\caption{A service call in Android Framework.} \label{fig:local}
\begin{lstlisting}
public class LocationManagerService {
  private UserManager mUserManager;
  void updateUserProfiles(int currentUserId) {
    List<UserInfo> profiles = mUserManager.getProfiles(currentUserId);
  }
  ...
}
public class UserManager {
  private final IUserManager mService;   
  public List<UserInfo> getProfiles(int uHandle) {
    mService.getProfiles(uHandle, false); 
  }
  ...
}
\end{lstlisting}
\aaf\aaf\af
\end{figure}



Service calls are frequently used among services. 
While inter-process service calls are made through the intricate Binder IPC mechanism,
intra-process calls are actually ordinary method calls. 
Figure~\ref{fig:local} shows an example, where the Location Manager service 
invokes \texttt{getProfiles} exposed by the User Manager service; both services
belong to the System Server process. 
The call at Line 4 leads to a service call at Line 11, which is a virtual
function call, whose dispatch relies on the runtime type of the object
pointed to by  \texttt{UserManager.mservice}.
Previous research relies on expert knowledge and specifies the dispatch targets 
manually to facilitate further analysis~\cite{kratos,edgeminer,pscout}, 
while \aeg makes use of the runtime type information provided by the execution context, 
and thus the call is handled as an ordinary virtual function call 
without requiring expert knowledge or manual effort. This is a concrete 
example illustrating the advantage of combining concrete and symbolic executions.

\subsubsection{Dealing with Handler and State Machine Calls} \label{sec:messagecall}

Two other important IPC mechanisms that are widely used by system services are 
Message Handler and State Machine calls. A handler sends and 
processes messages associated with a thread's message 
queue~\cite{handler}. When a new handler is created, it is bound to 
the message queue of the thread that creates it. From that point 
on, it will deliver messages to that message queue and execute them as 
they come out of the message queue. To deal with Message Handler calls, 
when \texttt{sendMessage(message)} is invoked, the invocation is replaced by
that of the corresponding Handler's \texttt{handleMessage(message)}.
The symbolic executor interposes  the \texttt{invokevirtual} 
instruction and enforces the replacement on the fly.

A State Machine can also send and process messages, which has states arranged 
hierarchically. A state is an instance of the \texttt{State} class, which implements 
\texttt{processMessage} for handling messages. A State Machine 
sends a message by invoking \texttt{sendMessage}. When a State 
Machine receives a message, the \emph{current state}'s 
\texttt{processMessage} is invoked. Therefore, a key step is to identify
the current state. To do it, the
field \texttt{mSmHandler} in the State Machine object, which is a reference 
to the state machine handler, is retrieved (note that when the State 
Machine object is migrated, all its fields are copied), and 
then used to migrate the state machine handler object. Next, two fields in the handler 
object, \texttt{mStateStack} and \texttt{mStateStackTopIndex}, are used to identify 
the current state (= \texttt{mStateStack[mStateStackTopIndex].state}). 
To handle messages sent by a State Machine in the symbolic executor, the invocation of
\texttt{mSmHandler.sendMessage(message)} is replaced by that of the current state's 
\texttt{processMessage(message)}. This way, we connect the senders and 
receivers for messages sent through State Machine.

\subsubsection{Handling Calls to Native Code} \label{sec:jni}
Part of Android Framework is implemented in native code, which
is invoked through the Java Native Interface (JNI) mechanism.
Different ways are adopted to handle JNI calls 
during symbolic execution. First, methods that return
the calling UID (\texttt{getCallingUid()}) and
the package name of the client app 
(\texttt{getPackageName()}) are modeled to 
return the corresponding information for the skeleton app constantly, and
the return values are set to be taint sources as aforementioned.
Second, the return values of other native methods that return 
app-specific information of the skeleton app are specified as symbolic inputs.
For example, many native methods declared
in the package \texttt{android.content.res} access application resources. 
Third, for native methods that do not have return values 
they are ignored; ignoring calls to external code has been 
used in many symbolic execution techniques~\cite{exe,spf}.

Finally, other calls to native methods are delegated
back to Android as remote procedure calls.   
The RPC client in the symbolic executor is built 
by extending jpf-nhandler~\cite{jpf-nhandler}.
While jpf-nhandler delegates native calls to a host JVM,
this client delegates them to an app running as the RPC server in 
a remote Android system (Figure~\ref{fig:arch}), which issues delegated 
native calls using reflection on demand. 
The \emph{GSON} library \cite{gson} is used for marshalling (and unmarshalling) method 
parameters and return values, which are transmitted between
the RPC server and the client via socket. Note that though
an Android system is used to execute native calls, the symbolic executor
is decoupled from it using the RPC mechanism.



\subsection{Other Aspects of Symbolic Execution} \label{sec:engine}

%


\subsubsection{Attributes for Symbolic Execution}
JPF supports attributes to be associated with program values including
locals, stack operands, and class/object fields in the heap. 
The framework makes use of attributes to store taints, reference flags, 
symbolic input attributes, and symbolic expressions. 


\subsubsection{Overriding Bytecode Interpretation}
JPF allows replacing or extending the interpretation of
bytecode instructions. Interpretation classes for instructions, such as
\texttt{getfield}, \texttt{getstatic}, and \texttt{aaload},
are overridden to specify symbolic inputs and migrate the 
execution context information.


\subsubsection{Intercepting Method Calls} \label{sec:implement-native}

JPF provides a mechanism called \emph{Model Java Interface} (MJI)
that intercepts method invocations for custom handling. \aeg makes use of MJI 
to intercept certain method calls (e.g., \texttt{getCallingUid}, \texttt{getPackageName}, and 
the \texttt{get} functions of various collection data structures),
and redirects them to our custom implementation of these functions. 
This mechanism is also used by the RPC client for intercepting
calls to native code.

\section{Evaluation} \label{sec:evaluation}

\subsection{Experiment Settings and Overview}

The experiments were performed on a machine with an Intel Core i7 4.0Ghz 
Quad Core processor and 32GB RAM running Linux kernel 3.13. 
Exploits were generated on Android Framework 5.0, and
verified using different versions of Android systems and settings.

We first present two case studies that demonstrate the applications of
\aeg. They show two typical scenarios of applying \aeg.
The first case study illustrates how static analysis and symbolic execution
are combined to find vulnerabilities and generate exploits. 
The second case study relies on the \aeg system only. 

Next, the reliability of the approach based on heap memory 
snapshots is investigated. We present exploit generation experiments 
based on snapshots captured at different times, and analyze
the results.

Finally, we compare symbolic execution used in \aeg against 
under-constrained symbolic execution (UCSE). Both can start symbolic
execution from system interface methods instead of the \texttt{main} function
to reach the code deep in the program, but \aeg makes
use of the execution context provided by concrete execution
to improve the precision and efficiency of 
symbolic execution.  



\subsection{Case Study 1: Exploiting Inconsistent Security Policy Enforcement (ISPE)}

\subsubsection{Background}
Android Framework utilizes a permission-based security model, 
which provides controlled access to various system resources.
However, a sensitive operation may be reached from different paths,
which may enforce security checks inconsistently. As a result,
an attacker with insufficient privilege may perform sensitive operations
by taking paths that lack security checks. Recently, static analysis 
combined with manual code inspection has been
applied to finding such inconsistent
security enforcement cases in Android Framework~\cite{kratos}.
The system, called \emph{Kratos}, first builds a call graph based on the Android Framework code. 
With the call graph, it finds all the execution paths that can reach
sensitive operations. Kratos then compares the paths pairwise to
identify paths that reach the same sensitive operation 
with inconsistent security checks enforced, and reports them as 
suspected ISPE vulnerabilities, as they violate the security property
that \emph{all paths should have consistent permissions for reaching
a given sensitive operation.} 


\subsubsection{Combined Approach for Bug Finding}
While static analysis is very scalable, it is well known that
the analysis results may be imprecise. In the case of finding
ISPE bugs, static analysis based on the reachability analysis 
may  report false positives, as some paths 
may be infeasible in real executions.
Currently, manual effort is used to scrutinize  
the code along each reported path, which is laborious and tedious;
moreover, it is difficult to verify the correctness of the manual inspection.

We propose to combine static analysis and symbolic execution
to find ISPE bugs. For each suspected vulnerability reported by
static analysis, \aeg (1) finds all
feasible paths that reach the sensitive operation, (2) gives permissions
needed for each feasible path (the needed permissions
are included in each path condition), (3) verifies permission consistency 
among the feasible paths, and (4) generates inputs that
exercise the feasible paths to verify suspected vulnerabilities. 
It thus demonstrates the applications of \aeg comprehensively.
All the steps have been performed automatically, 
in contrast with previous work that relies on tedious and error-prone manual inspection.
In addition, zero false positives are guaranteed as all suspected vulnerabilities
are validated by the generated inputs.


\begin{table*}[t]
\caption{List of vulnerabilities. (\emph{LMS}, \emph{PSB}, 
\emph{TSI}, \emph{PIM}, \emph{WMS}, \emph{AMS}, \emph{WSI}, \emph{NS}, and 
\emph{ASS} represent LocationManagerService, PhoneStateBroadcaster, 
TelecomServiceImpl, PhoneInterfaceManager, WindowManagerService, ActivityManagerService,  
WifiServiceImpl, NsdService, and ActivityStackSupervisor, respectively.)} \label{all} 

\centering
{\small \footnotesize   
\scalebox{0.88}{
\begin{tabular}{c|c|c|c|c|c|c|c|c|c|c||c}
\hline
\multirow{3}{*}{No.} & \multirow{2}{*}{Vulnerability} &  \multirow{3}{*}{Entrypoint(s)} & \multicolumn{2}{c|}{\# of} & \multicolumn{2}{c|}{\# of} & \# of  & \# of  & \multirow{1}{*}{Sym.} & \multirow{3}{*}{Exploitable?} & \multirow{2}{*}{Consistent with} \\
 & \multirow{2}{*}{description} &   & \multicolumn{2}{c|}{migrated classes} & \multicolumn{2}{c|}{migrated objects} & all &  legal &  \multirow{1}{*}{exe.} &  &  \multirow{2}{*}{Kratos?} \\
\cline{4-7} 
 & &  &  min & max & min & max & sets & sets &  \multirow{1}{*}{time}  & & \\

\hhline{=|=|=|=|=|=|=|=|=|=|=||=}

\multirow{3}{*}{1} & Access  & \multirow{2}{*}{LMS.getAllProviders()}  & \multirow{2}{*}{55} & \multirow{2}{*}{55} & \multirow{2}{*}{4} & \multirow{2}{*}{4}  & \multirow{2}{*}{---} & \multirow{2}{*}{---} & \multirow{2}{*}{14s} & \multirow{3}{*}{\ding{51}} & \multirow{3}{*}{\ding{51}}  \\ 
& installed providers &  \multirow{2}{*}{LMS.getProviders(Criteria,boolean)} & \multirow{2}{*}{77} & \multirow{2}{*}{93} & \multirow{2}{*}{14} & \multirow{2}{*}{42}  & \multirow{2}{*}{130} & \multirow{2}{*}{130} & \multirow{2}{*}{1m 28s} & &\\ 
& with insuf. privilege & & &  & & & & &  & & \\ 
\hline

\multirow{3}{*}{2} & Read  & TSI.getCallState() & 48 & 48 & 3 & 3 &  --- & --- & 5s & \multirow{3}{*}{\ding{51}}  & \multirow{3}{*}{\ding{51}} \\ 
& phone state &  TSI.isInCall() & 62 & 69 & 17 & 20 & 1  & 1 & 23s & &\\ 
& with insuf. privilege  & TSI.isRinging() & 60 & 65 & 16 & 18 & 1  & 1 & 21s & & \\ 
\hline

\multirow{3}{*}{3} & End  & \multirow{2}{*}{TSI.endCall()} & \multirow{2}{*}{81} & \multirow{2}{*}{83} & \multirow{2}{*}{21} & \multirow{2}{*}{24} &  \multirow{2}{*}{1} & \multirow{2}{*}{1} & \multirow{2}{*}{18s} & \multirow{3}{*}{\ding{51}} & \multirow{3}{*}{\ding{51}} \\ 
& phone calls & \multirow{2}{*}{PIM.endCall()} & \multirow{2}{*}{80} & \multirow{2}{*}{85} & \multirow{2}{*}{23} & \multirow{2}{*}{26} & \multirow{2}{*}{1} & \multirow{2}{*}{1} & \multirow{2}{*}{24s}  & & \\ 
& with insuf. privilege  &  &  &  & & & &  & & & \\ 
\hline    

\multirow{3}{*}{4} & Close  &  \multirow{2}{*}{WMS.closeSystemDialogs(String)} & \multirow{2}{*}{57} & \multirow{2}{*}{57} & \multirow{2}{*}{6} & \multirow{2}{*}{6} & \multirow{2}{*}{---} & \multirow{2}{*}{---} & \multirow{2}{*}{11s} & \multirow{3}{*}{\ding{51}}  & \multirow{3}{*}{\ding{51}} \\ 
& system dialogs  &  \multirow{2}{*}{AMS.closeSystemDialogs(String)} & \multirow{2}{*}{63} & \multirow{2}{*}{67} & \multirow{2}{*}{11} & \multirow{2}{*}{15} & \multirow{2}{*}{2} & \multirow{2}{*}{2} & \multirow{2}{*}{37s} & & \\ 
& with insuf. privilege  &  &  & & & & &  & & & \\ 

\hline 
\multirow{3}{*}{5} & Set up HTTP proxy & \multirow{2}{*}{WSI.addOrUpdateNetwork()} & \multirow{2}{*}{67} & \multirow{2}{*}{122} & \multirow{2}{*}{23} & \multirow{2}{*}{52} &  \multirow{2}{*}{18} & \multirow{2}{*}{18} & \multirow{2}{*}{1m 06s} & \multirow{3}{*}{\ding{51}} & \multirow{3}{*}{\ding{55}}   \\ 
& working in PAC mode &  \multirow{2}{*}{WSI.getWifiServiceMessenger()} & \multirow{2}{*}{65} & \multirow{2}{*}{84} & \multirow{2}{*}{21} & \multirow{2}{*}{24} & \multirow{2}{*}{1} & \multirow{2}{*}{1} & \multirow{2}{*}{43s} & & \\ 
& with insuf. privilege  &  &  & & & & &  & & & \\ 

\hline 
\multirow{3}{*}{6} & Enable/Disable &  \multirow{2}{*}{NS.setEnabled(boolean)} & \multirow{2}{*}{75} & \multirow{2}{*}{114} & \multirow{2}{*}{28} & \multirow{2}{*}{53}  & \multirow{2}{*}{1} & \multirow{2}{*}{1} & \multirow{2}{*}{37s} & \multirow{3}{*}{\ding{51}} & \multirow{3}{*}{\ding{55}}  \\ 
& mDNS daemon & \multirow{2}{*}{NS.getMessenger()} & \multirow{2}{*}{80} & \multirow{2}{*}{81} & \multirow{2}{*}{11} & \multirow{2}{*}{14} & \multirow{2}{*}{1} & \multirow{2}{*}{1} & \multirow{2}{*}{45s} & & \\ 
& with insuf. privilege  &  &  & &  &  & & & & \\ 



 
\hhline{=|=|=|=|=|=|=|=|=|=|=||=}
\multirow{3}{*}{7} & \multirow{2}{*}{Task hijacking} & \multirow{2}{*}{ASS.startActivityUncheckedLocked()}  & \multirow{2}{*}{324} & \multirow{2}{*}{387} & \multirow{2}{*}{136} & \multirow{2}{*}{182} &  \multirow{2}{*}{2,020} & \multirow{2}{*}{810} & \multirow{2}{*}{14m 33s}  & \multirow{2}{*}{\ding{51}}  &  \multirow{2}{*}{N/A}  \\ 
& &  &  & & & & & & & & \\ 

\hline
\end{tabular}
}
}

\end{table*}

Table~\ref{all} summarizes the experiment results (the vulnerability shown 
in the last row is discussed in  Case Study 2).
For each vulnerability, the table lists 
the vulnerability description,
entrypoint(s), the min/max number of migrated classes among different 
paths, the min/max number of migrated objects among different paths, the number 
of sets of concrete values generated (``---'' means it can be exploited unconditionally), 
the number of sets that can be used to generate exploits, the symbolic execution 
time, whether the suspected vulnerability is really exploitable,  and 
whether the results are consistent with those of Kratos. 

Given an entrypoint method, there are may be multiple paths that reach
the sensitive operation, and the classes and objects involved in
the paths may vary, as illustrated by the min/max number of
migrated classes and objects. Note that when migrating a class, 
all its super classes are also migrated, which is the reason
the number of migrated classes is greater than that of objects.
In the majority of the cases, the symbolic execution of an entrypoint
method is finished within less one minute.
All the cases are verified using the inputs generated by \aeg,
showing they are exploitable. 

\textbf{New findings.} It is notable that some of
our results are inconsistent with those of Kratos.
First,  for the fifth vulnerability in Table~\ref{all}, Kratos
reports that it does not exist in Android Framework 5.0, while
\aeg shows that it still exist (i.e., different permissions are required
by the two system interface methods for reaching the sensitive resource) and the 
result is verified using inputs generated by \aeg. 
Second, for the sixth vulnerability in Table~\ref{all}, 
 Kratos reports only one permission 
\texttt{CONNECTIVITY\_INTERNAL} for invoking \texttt{NsdService.setEnabled}, while
\aeg\ reports two permissions, 
\texttt{CONNECTIVITY\_INTERNAL} and \texttt{WRITE\_SETTINGS}.
The more thorough and accurate results demonstrate the advantages
of the hybrid approach.

\subsubsection{A Detailed Example}
As an example, we describe in detail how the combined approach was applied to finding the
first vulnerability in Table~\ref{all}. First, the static analysis based on 
path reachability and pairwise path comparison finds that  both
\texttt{getProviders(Criteria, boolean)}
and \texttt{getAllProviders()} (in the \texttt{LocationManagerService} class)
have paths reaching the same sensitive operation that returns
the names of the installed GPS providers, and the two paths can be executed
with inconsistent permissions; thus, it is a suspected 
vulnerability.
Next, \aeg is applied to validating it automatically. Specifically,
after the Android system is initialized and the skeleton app is launched, 
a heap memory snapshot
of the System Server process is captured, and provides execution
context for symbolic execution, and symbolic execution starts from
the two service interface methods respectively. 

\begin{figure} [t]
\centering
\aaf \af
\hspace*{-12pt}
\includegraphics[scale=0.65]{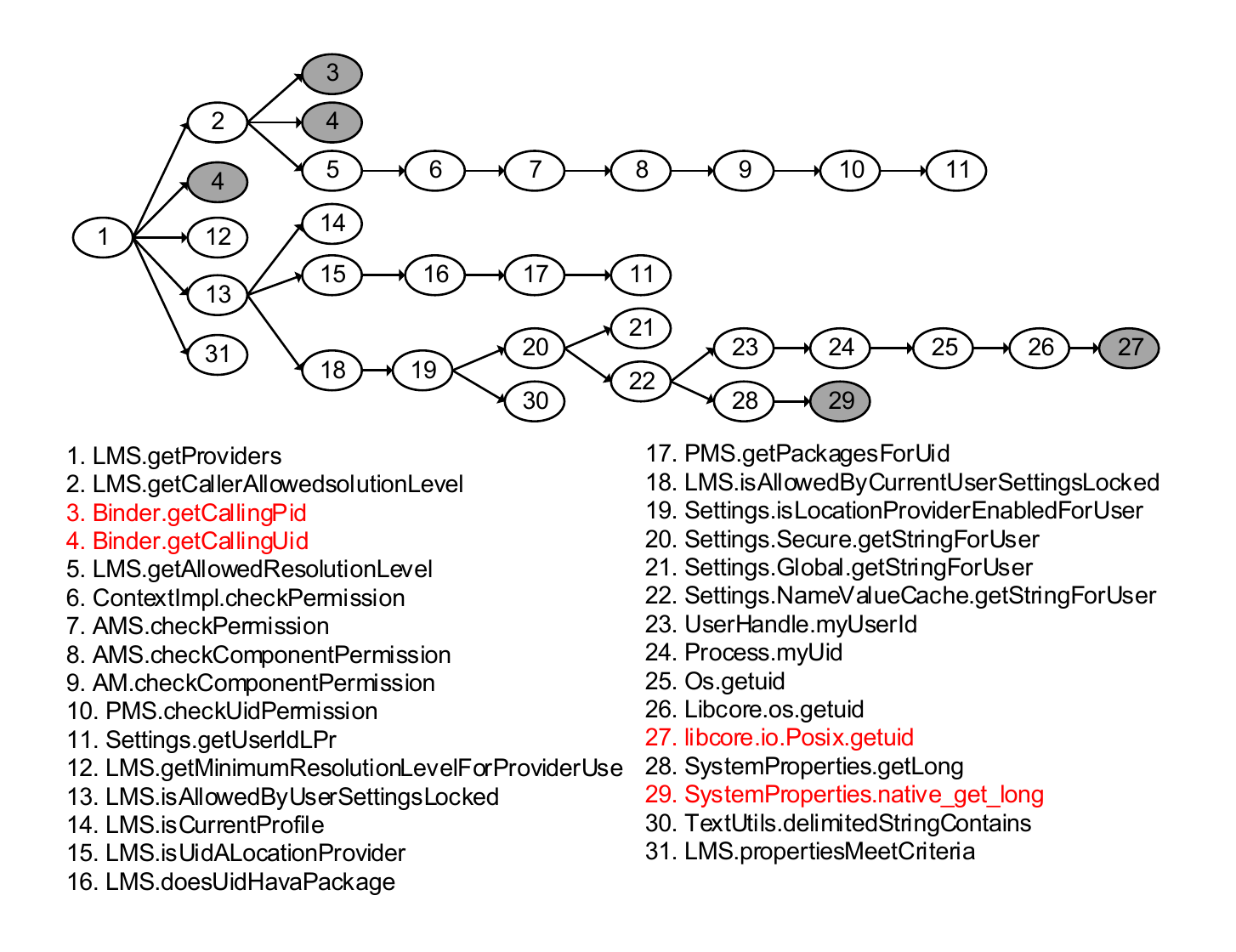}
\aaf \aaf \aaf\aaf 
\caption{Sub-call graph rooted at \texttt{getProviders(Criteria, boolean)}. 
(\emph{LMS}, \emph{AM},
\emph{AMS}, and \emph{PMS} represent LocationManagerService, 
ActivityManager, ActivityManagerService, and PackageManagerService,
respectively. 
The grey nodes denote native methods.)
} 
\aaf 
\label{getprovider-call}
\end{figure}

\textbf{Entrypoint 1}: \texttt{getProviders(Criteria, boolean)}. 
Figure~\ref{getprovider-call} shows the sub-call graph rooted at 
this entrypoint with collection
and string operations omitted. It leads to invocation of multiple methods 
of other services, e.g.,
\texttt{ActivityManagerService} and \texttt{PackageManagerService}.
The services run in the same process, so are handled as ordinary
method calls using the runtime type information in the execution context.

Four native methods are involved: \texttt{getCallingUid}, \texttt{getCallingPid},
\texttt{getuid}, and \texttt{native\_get\_long}. Calls to these methods are 
intercepted using MJI, and are redirected to our handlers of these methods. 
The first two return the UID and PID of the client app, respectively,
and \texttt{getuid} returns the UID $=1000$, which is the UID of 
the System Server process. The call to \texttt{native\_get\_long}
is delegated back to the Android system through RPC.


\begin{figure}
\caption{Examples of concrete input values generated in case study 1.} \label{concrete}
\begin{lstlisting}[language=java,numbers=none]
(mUserIds.array[54].grantedPermissions.backingMap.table[836059052 & (length_SYM - 1].key_SYM == permission.ACCESS_FINE_LOCATION) &&
(criteria != null) && 
(criteria.mHorizontalAccuracy == 2) &&
(criteria.mPowerRequirement == 0) &&
(criteria.mAltitudeRequired == false) &&
(criteria.mSpeedRequired == false) &&
(criteria.mBearingRequired == false) &&
(criteria.mCostAllowed == false)
(enabledOnly = false) && 
//output: ["gps"]
\end{lstlisting}
\aaf
\begin{lstlisting}[language=java,numbers=none] 
(mUserIds.array[54].grantedPermissions.backingMap.table[836059052 & (length_SYM - 1].key_SYM == permission.ACCESS_FINE_LOCATION) &&
(criteria == null) &&
(enabledOnly = true) && 
//output: ["gps", "passive"]
\end{lstlisting}
\aaf
\aaf\af
\end{figure}

The variable \texttt{mUserIds.array[54]} is identified as a symbolic input 
through the tainting during symbolic execution.
Figure~\ref{concrete} shows several examples of the generated concrete values.
Take the first set as an example; it provides clear information for
building an app that exercises the paths  in terms of configuring
the app (i.e., requiring the 
\texttt{ACCESS\_FINE\_LOCATION} permission) and preparing the parameter values
(i.e., \texttt{criteria} and \texttt{enabledOnly}) for invoking 
the entrypoint method. 

\textbf{Entrypoint 2}: \texttt{getAllProviders()}. The generated
path condition is constantly true, which means this method
can be invoked with no permissions needed. 

As the needed permissions required by the two entrypoints differ, 
it is identified as an ISPE vulnerability.
We then checked the reliability of the exploits. 
60 emulators with different device types, Android framework versions (4.3, 4.4, and 5.0), 
and CPU/ABI configurations were used to check whether the completed
apps could access the targeted sensitive resource. The experiments
show that the apps are effective on all the emulators, as
the different configurations among the emulators do not affect
the invocation and execution of the system interface methods. 
It demonstrates the reliability of the exploits.  

\textbf{Summary.} Compared to previous work that relies on
enormous and error-prone manual inspection,
the combined approach of static analysis and symbolic execution
eliminates the need for manual work and guarantees zero false positives.
It is potential to apply this approach to finding other types of vulnerabilities
in Android Framework. 

\subsection{Case Study 2: Constructing Task Hijacking Attacks}

\subsubsection{Background}
The Activity Manager Service (AMS)  allows
activities of different apps to reside in the same \emph{task}, which
is a collection of activities that users interact with when performing 
a certain job. The activities in a given task are arranged in a \emph{back stack}, 
pushed in the order they were opened; users can navigate back using
the ``Back'' button. This feature can be exploited by a malicious app if its activities
are manipulated to reside side by side with the victim apps in the same task and hijack 
the user sessions of the victim apps. 

This is a design flaw rather than a program bug, and can be exploited
to implement UI spoofing, denial-of-service, and user monitoring attacks~\cite{task}.
For example, a malicious app may start a malicious activity that impersonates 
the victim activity, and the UI spoofing attack succeeds if the 
fake activity resides in the same back stack as the target victim activity,
and the user may mistake the fake malicious activity for the victim one.  
The security property here is that \emph{the malicious activity should
not reside in the same back stack as the target victim activity.}

This case illustrates unique characteristics of generating exploits that
take advantage of Android Framework vulnerabilities: while the design 
flaw is due to Android Framework, the victim entity is not the framework but another app, and the
malicious ``input'' is not a simple string input but a separate app.
It is very different from exploit generation targeting a vulnerable
executable, which typically involves a single entity (the vulnerable executable)
and the attack input is usually in the form of a string.

\subsubsection{Bug Finding}
We use the \texttt{EditEventActivity} activity of the
\texttt{com.android.calendar} app as an example victim activity.
In the skeleton app, the main activity of the skeleton app starts 
the malicious activity, denoted by $M$. 
The goal of the attack is that $M$, when it is started, will reside 
in the same task as the victim activity. 
A bug is identified if such attacks against the victim activity is feasible. 
We capture the heap memory snapshot when 
 the victim app and the skeleton app are started and the main
activity of the skeleton app is ready to start the malicious activity.

\begin{figure}[t]
\centering
\aaf \af
\hspace*{-9pt}
\includegraphics[scale=0.75]{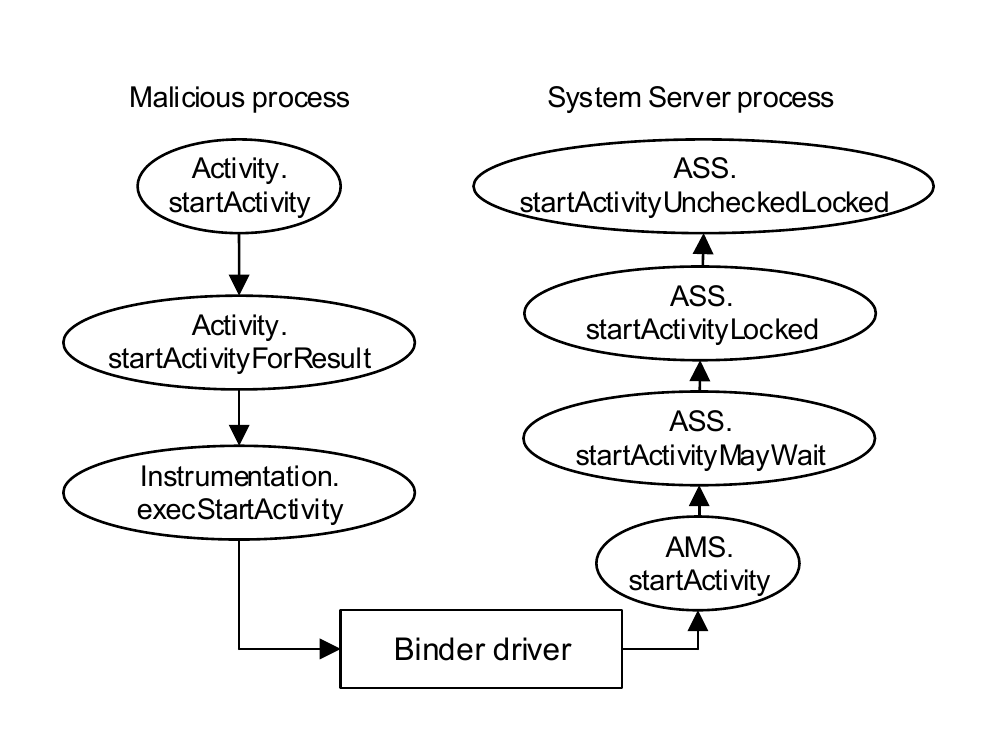}
\captionsetup{font={footnotesize}}
\aaf \aaf \aaf
\caption{The rough procedure of starting an activity. (ASS represents ActivityStackSupervisor.)} 
\af
\label{startactivity}
\end{figure} 

\begin{figure}
\caption{Function \texttt{startActivityUncheckedLocked}.} \label{fig:function}
\begin{lstlisting}[language=java,numbers=none]
final int startActivityUncheckedLocked(ActivityRecord r, ActivityRecord sourceRecord, IVoiceInteractionSession voiceSession, IVoiceInteractor voiceInteractor, int startFlags, boolean doResume, Bundle options, TaskRecord inTask) {...}
\end{lstlisting}
\aaf\af
\end{figure}

Figure~\ref{startactivity} shows the rough procedure of starting 
the malicious activity. The API \texttt{Activity.startActivity(Intent, Bundle)} 
is invoked with the parameters specifying
the activity to be started, i.e., $M$. The invocation leads
to a service request to be handled by the service interface method 
\texttt{startActivity} in AMS. The operations of selecting the task hosting 
the new activity are performed in \texttt{startActivityUncheckedLocked}, which
has eight parameters as shown in Figure~\ref{fig:function}.
The first parameter \texttt{r} is an \texttt{ActivityRecord} instance storing the 
information of $M$, while the second storing that of the caller activity. 
The description of other parameters is omitted. They are set to symbolic inputs.

The constraint indicating that \emph{the task selected for $M$ is exactly 
the one hosting the victim activity} is added to each of the path conditions
when it is to be resolved. A feasible path
is found if the path condition is resolvable. 

\subsubsection{Exploit Generation}

\begin{figure}
\caption{Examples of concrete input values generated in case study 2.} \label{concrete2}
\begin{lstlisting}[language=java,numbers=none]
// Illegal concrete values 
(r.intent.mFlags == 0x11000000) && 
(r.launchMode == LAUNCH_SINGLE_TOP) &&
(r.mLaunchTaskBehind == true) && 
(options == null) &&
(r.info.documentLaunchMode == 0) &&
(r.info.targetActivity == null) && 
(r.resultTo == null) &&
(r.taskAffinity != "android.task.calendar") && 
(r.intent.mComponent.mClass == "com.android.calendar.EditEventActivity") &&
(r.intent.mComponent.mPackage == "com.android.calendar")
\end{lstlisting}
\aaf
\begin{lstlisting}[language=java,numbers=none]
// Legal concrete values
(r.intent.mFlags == 0x10080000) && 
(r.launchMode == LAUNCH_SINGLE_TASK) &&
(r.mLaunchTaskBehind == true) && 
(options == null) &&
(r.resultTo == null) &&
(r.info.documentLaunchMode == 0) &&
(r.info.targetActivity == null) && 
(r.taskAffinity == "android.task.calendar") 
\end{lstlisting}
\aaf
\end{figure} 

The symbolic execution generated 2,020 sets of concrete input values, among which
some contain illegal concrete values, e.g., due to requiring the 
malicious activity's package and activity names to be equal to those of the victim 
activity. Simple scripts were written to filter out illegal concrete values,
the number of which is 1,210 sets totally. Figure~\ref{concrete2} (upper part) shows an example of illegal 
concrete values; it is illegal because its package name is duplicate with that of the victim app, but
Android requires that each app should have a unique package name.

Figure~\ref{concrete2} (lower part) shows an 
example of the rest 810 sets of legal concrete values. 
In this example, \emph{r.intent.mFlags} and \emph{options} guide how to 
set the input parameters of \texttt{startActivity} for starting the malicious 
activity; others instruct how to configure the malicious activity; for example, 
\texttt{r.launchMode} is mapped to the \texttt{android:launchMode} in the manifest file. 
Figure~\ref{build} shows the exploit according to the set of concrete values.
When users click the app icon of the malicious app in the home screen, the main 
activity will be started and it is coded to 
call \texttt{startActivity} to start the malicious activity, which
will reside side by side with the victim activity in the same task.

\begin{figure}
\caption{Exploit generated from the concrete input values.}
\begin{lstlisting}[language=java,numbers=none]
// Snippet of AndroidManifest.xml
<activity android:name=".maliciousActivity" 
          android:launchMode="singleTask" 
          android:taskAffinity="android.task.calendar"
          android:documentLaunchMode="none" />

// The main activity starts the malicious activity
public void onCreate(Bundle savedInstanceState) {
  ...
  Intent i = new Intent(this, maliciousActivity.class);
  intent.setFlags(0x10080000);
  startActivity(i, null);
}
\end{lstlisting}\label{build}
\end{figure}

\begin{table}[t]
\small
\caption{Effectiveness of the generated exploits.}\label{exploits2}
\centering
{
\begin{tabular}{c | c c c c c c} 
\hline
Android version & 4.0 & 4.1 & 4.2 & 4.3 & 4.4 & 5.0 \\
\hline
\# of effective exploits & 434 & 674 & 674 & 674 & 702 & 810 \\
\hline
\end{tabular}
}
\end{table}

We then examined whether the exploits generated on Android 5.0 were  
 effective on other versions of Android systems. 
Table~\ref{exploits2} lists the results, which show that the effectiveness of the
exploits are affected by the versions of Android systems. Further investigation
has revealed that the difference is mainly caused by code changes.  
For example, the new exploiting condition \texttt{FLAG\_ACTIVITY\_NEW\_DOCUMENT} is not
introduced until Android 5.0 (discussed below);
the API \texttt{startActivity(Intent, Bundle)} 
is not included in version 4.0, and thus only exploits with \texttt{options == null}
can be used for invoking  \texttt{startActivity(Intent)}.

\begin{figure} [t]
\caption{Exploiting condition.} \label{fig:condition}
\begin{lstlisting}[language=java,numbers=none]
(((((r.intent.mFlags & 0x7F7FFFFF) | 0x10000000) | 0x8000000) | 0x10000000) & 0x80000) != 0x80000
\end{lstlisting}
\af \af
\end{figure}

\textbf{Newly discovered exploiting condition.} The path conditions generated
from symbolic execution reveal a new exploiting condition, as shown in 
Figure~\ref{fig:condition}, that was not reported in previous work~\cite{task}. 
Here, 0x80000 represents the flag \texttt{FLAG\_ACTIVITY\_NEW\_DOCUMENT}, 
which is introduced since Android 5.0, and the seemingly complex condition 
simply means the corresponding bit in the bitflags $r.intent.mFlags$ is 0. 
Compared to previous work that relies on 
\latin{ad hoc} manual effort for discovering the exploiting
conditions, \aeg\ finds them in a systematic and automatic way.


\subsection{Consistency of Exploits Generated with Different Snapshots}

We then investigated whether snapshots captured at different times 
affected exploit generation. After the system is initialized, 20 snapshots were captured at intervals of
5 minutes on Android 5.0 with random user interactions during the intervals. 
For each vulnerability listed in Table~\ref{all},
symbolic execution was performed with each of the 20 snapshots providing
the execution context. The results show that, for each vulnerability,
the same sets of path conditions were generated with different snapshots, 
which means that the resulting exploits with the different
snapshots are consistent.

There are several reasons that explain the consistency of exploits. 
First, if a malicious app 
does not rely on other apps to exploit a vulnerability (e.g., inconsistent security policy 
enforcement), access control is enforced in Android Framework to 
make sure the information of other apps is not accessed. 
Thus, the configurations and statuses of other apps do not affect the
path exploration. 
On the other hand, for exploits that rely on the statuses of other apps
(e.g., the victim app in task hijacking attacks), the path exploration probably depends on
the statuses of one or more apps. During symbolic execution, reasonable
setting up is established consistently; for example, the victim activity should already be started
in the task hijacking case prior to capturing snapshots. The results show that
an attack succeeds as long as the same statuses recur. 

Finally, the values of non-app-specific variables do not affect path exploration
for the vulnerabilities we examined. For example, in the case of inconsistent
security policy enforcement for accessing the names of installed providers, 
the path exploration does not depend on the concrete values of the related 
non-app-specific variable (i.e., \texttt{LocationManagerService.mProviders}),
although different provider names may be returned by the service calls if different providers are installed.


\subsection{Comparison with Under-constrained Symbolic Execution (UCSE)}

Both \aeg\ and UCSE are able to start symbolic execution at any service interface
method of Android Framework. The major difference between the two is that 
UCSE does not have the type and value information about the inputs, 
while \aeg\ obtains the information from the execution context provided by the 
concrete execution. 

The first issue of applying UCSE to symbolic execution of Android Framework
is that virtual function calls are frequently used, but the runtime types of the 
receiver objects are unknown. 
UCSE constructs the receiver objects 
either using lazy initialization based on the type hierarchy or relying on
manual specifications, which either explores spurious paths or requires much manual effort. 

The second issues is that input variables which are treated as concrete inputs in \aeg\ are 
treated as symbolic inputs in UCSE. 
UCSE handles such symbolic inputs using lazy initialization, which causes the
following problems: (1) loops that iterate through collection data structures are
unbounded, and (2) the generated concrete values are unrealistic.

\begin{table} 
\centering
\caption{Execution time of UCSE and \aeg.}\label{exetime}
{\small 
\begin{tabular}{@{~}c@{~}|@{~}c@{~}|@{~}c@{~}}
\hline
Vulnerability & UCSE & \aeg \\
\hline
ISPE in accessing GPS providers   & Out-of-mem & 1m 42s \\ 
Task hijacking & Out-of-mem & 14m 33s \\ 
\hline
\end{tabular}
}
\aaf\af
\end{table}

We tried to perform UCSE of Android Framework using Java PathFinder, which kept crashing
when it was applied directly. We spent a lot of time and tedious effort modifying
the framework code (e.g., adding the type information about objects pointed to by references
to assist dynamic dispatching) to make the symbolic execution possible.
We thus only modified the code with respect to  the ISPE vulnerability
of accessing the GPS provider list and the task hijacking vulnerability. 
Table~\ref{exetime} shows the execution time applying the two techniques. 
Due to path explosion UCSE in both cases ran out of memory. 

Therefore, path exploration without precise information of the execution
context causes many problems, such as requiring tedious 
manual effort, generating unrealistic outputs, and exploring spurious paths. 
\aeg\ resolves the problems by migrating
the execution context from the concrete execution world to symbolic execution.

\section{Related Work} \label{sec:related}
\noindent \textbf{Mixing Concrete/Symbolic Execution.} DART is the first 
concolic testing tool that uses symbolic analysis in 
concert with concrete execution to improve coverage of random testing~\cite{dart}. 
It runs the tested unit code on random inputs and symbolically gathers constraints at decision 
points that use input values; then, it negates one of these symbolic constraints 
to generate the next test case. 
EGT~\cite{egt}, EXE~\cite{exe} and KLEE~\cite{klee} 
execute external code concretely by using one of the possible
concrete values of the symbolic operands.
S$^2$E introduces selective symbolic execution, which allows
a program's paths to be explored without having to model its surrounding environment~\cite{s2e}.
These techniques usually take advantage of concrete execution to simplify 
complex symbolic constraints and execute external code, while \aeg\
makes use of concrete execution to set up the execution
context for symbolic execution.

\noindent \textbf{Switching Concrete Execution to Symbolic Execution.}
Symbolic PathFinder (SPF)  begins with 
concrete execution and can switch to symbolic
execution at any point in the program~\cite{combined-spf}.
\aeg\ allocates concrete execution 
and symbolic execution to two decoupled systems,  
so that the two systems can evolve independently. 
As SPF aims at generating unit test cases, it simply specifies function parameters as 
symbolic inputs, while \aeg\ finds out
variables derived from the malicious app and uses them as symbolic inputs.

\noindent \textbf{Bug Finding.}
Fuzzing and symbolic execution have been applied to 
checking the existence of bugs. For example,
Miller et al.\ proposed a blackbox fuzzing technique that 
sends unstructured random inputs to an application program and considers a 
failure to be a crash or hang~\cite{miller}. It is mainly used
to reveal input validation bugs~\cite{mulliner2009injecting, inputvalid}.  SAGE
is a bug finding system~\cite{whitebox},
which leverages the technique described in DART~\cite{dart}. 
SAGE has demonstrated symbolic execution can be very useful for
bug finding. Through symbolic execution of Android Framework, \aeg shows its
effectiveness for bug finding as well.


\noindent \textbf{Exploit Generation.}
Automatic patch-based exploit generation (APEG) generates exploits based on
information in patches~\cite{apeg}. 
Compared to APEG, AEG does not require access to patches.
Both APEG and AEG target stand-alone native executables for exploit generation,
while \aeg\ considers exploit generation in an environment that manages all executables running on it.
Many unique challenges not seen in stand-alone executables have to be addressed by \aeg.


\noindent \textbf{Symbolic Execution of Android Apps.}
There has been a lot of work that leverages symbolic execution for
testing Android apps.
For example, Jensen et al.\ proposed to use concolic execution to build summaries of
the individual event handlers and then generate event sequences backward,
in order to  find event sequences that reach a given target line of code in the Android app~\cite{jensen}.
SIG-Droid combines program analysis techniques with symbolic execution
to generate event sequences as well as input values~\cite{sigdroid}.
All  use symbolic execution to exercise application code. 
To our knowledge, our system is the first one that supports
symbolic execution of Android Framework.

\section{Conclusions} \label{sec:conclusions}
We have introduced the first system for automatic generation of exploits
that take advantage of Android Framework vulnerabilities.
To avoid state space explosion due to the complex initialization,
concrete execution is used for the initialization phase,
providing execution context to symbolic execution.
Among the large number of variables in execution context, slim tainting 
tracks characteristic access patterns to identify variables
derived from the malicious apps as symbolic inputs.
In order to decouple the implementation of \aeg from Android,
the execution context provided by concrete execution is migrated
from the Android ART VM to a Java VM. 
We have implemented the system and evaluated it. The evaluation
shows that \aeg is very effective in both bug finding and exploit generation.
\aeg is also the first system that enables symbolic execution
of Android Framework.
Given that symbolic execution has proven to be a very useful technique,
we plan to apply \aeg to other purposes in future work, such as automatic
API specification generation, fine-grained malware analysis, and testing.




\bibliographystyle{abbrv}
\bibliography{refer}  

\begin{thebibliography}{10}

\bibitem{toxic}
{Android fragmentation turning devices into a toxic hellstew of
  vulnerabilities}.
\newblock
  \url{http://www.zdnet.com/article/android-fragmentation-turning-devices-into-a-toxic-hellstew-of-vulnerabilities/}.

\bibitem{aidl}
{Android interface definition language (AIDL)}.
\newblock \url{https://developer.android.com/guide/components/aidl.html}.

\bibitem{manifest}
{App Manifest}.
\newblock
  \url{https://developer.android.com/guide/topics/manifest/manifest-intro.html}.

\bibitem{flowdroid}
S.~Arzt, S.~Rasthofer, C.~Fritz, E.~Bodden, A.~Bartel, J.~Klein, Y.~L. Traon,
  D.~Octeau, and P.~McDaniel.
\newblock {FlowDroid}: Precise context, flow, field, object-sensitive and
  lifecycle-aware taint analysis for {Android} apps.
\newblock In {\em PLDI}, 2014.

\bibitem{pscout}
K.~W.~Y. Au, Y.~F. Zhou, Z.~Huang, and D.~Lie.
\newblock {PScout}: analyzing the android permission specification.
\newblock In {\em CCS}, 2012.

\bibitem{aeg}
T.~Avgerinos, S.~K. Cha, B.~L.~T. Hao, and D.~Brumley.
\newblock {AEG}: automatic exploit generation.
\newblock In {\em Communications of the ACM}, 2014.

\bibitem{apeg}
D.~Brumley, P.~Poosankam, D.~Song, and J.~Zheng.
\newblock Automatic patch-based exploit generation is possible: Techniques and
  implications.
\newblock In {\em USENIX Security}, 2008.

\bibitem{klee}
C.~Cadar, D.~Dunbar, and D.~R. Engler.
\newblock {KLEE}: unassisted and automatic generation of high-coverage tests
  for complex systems programs.
\newblock In {\em OSDI}, 2008.

\bibitem{egt}
C.~Cadar and D.~Engler.
\newblock Execution generated test cases: how to make systems code crash
  itself.
\newblock {\em Model Checking Software}, 2005.

\bibitem{exe}
C.~Cadar, V.~Ganesh, P.~M. Pawlowski, D.~L. Dill, and D.~R. Engler.
\newblock {EXE}: {automatically} generating inputs of death.
\newblock In {\em CCS}, 2006.

\bibitem{inputvalid}
C.~Cao, N.~Gao, P.~Liu, and J.~Xiang.
\newblock Towards analyzing the input validation vulnerabilities associated
  with android system services.
\newblock In {\em ACSAC}, 2015.

\bibitem{edgeminer}
Y.~Cao, Y.~Fratantonio, A.~Bianchi, M.~Egele, C.~Kruegel, G.~Vigna, and
  Y.~Chen.
\newblock {EdgeMiner}: automatically detecting implicit control flow
  transitions through the android framework.
\newblock In {\em NDSS}, 2015.

\bibitem{s2e}
V.~Chipounov, V.~Kuznetsov, and G.~Candea.
\newblock {S2E}: a platform for in-vivo multi-path analysis of software
  systems.
\newblock In {\em ASPLOS}, 2011.

\bibitem{Costa:2007:bouncer}
M.~Costa, M.~Castro, L.~Zhou, L.~Zhang, and M.~Peinado.
\newblock Bouncer: securing software by blocking bad input.
\newblock In {\em SOSP}, 2007.

\bibitem{CVE-2015-6628}
{CVE-2015-6628}.
\newblock \url{https://www.cvedetails.com/cve/CVE-2015-6628/}.

\bibitem{CVE-2016-2496}
{CVE-2016-2496}.
\newblock \url{https://www.cvedetails.com/cve/CVE-2016-2496/}.

\bibitem{CVE-2016-3750}
{CVE-2016-3750}.
\newblock \url{https://www.cvedetails.com/cve/CVE-2016-3750/}.

\bibitem{CVE-2016-3759}
{CVE-2016-3759}.
\newblock \url{https://www.cvedetails.com/cve/CVE-2016-3759/}.

\bibitem{taintdroid}
W.~Enck, P.~Gilbert, B.-G. Chun, L.~P. Cox, J.~Jung, P.~McDaniel, and A.~N.
  Sheth.
\newblock {TaintDroid}: An information-flow tracking system for realtime
  privacy monitoring on smartphones.
\newblock In {\em OSDI}, 2010.

\bibitem{under1}
D.~Engler and D.~Dunbar.
\newblock Under-constrained execution: marking automatic code destruction easy
  and scalable.
\newblock In {\em ISSTA}, 2007.

\bibitem{dart}
P.~Godefroid, N.~Klarlund, and K.~Sen.
\newblock {DART}: directed automated random testing.
\newblock In {\em PLDI}, 2005.

\bibitem{whitebox}
P.~Godefroid, M.~Y. Levin, and D.~Molnar.
\newblock Automated whitebox fuzz testing.
\newblock In {\em NDSS}, 2008.

\bibitem{framework}
Google.
\newblock {Android Interfaces and Architecture}.
\newblock \url{https://source.android.com/devices/}.

\bibitem{gson}
{GSON}.
\newblock \url{https://sites.google.com/site/gson/Home}.

\bibitem{handler}
{Handler}.
\newblock
  \url{https://developer.android.com/reference/android/os/Handler.html}.

\bibitem{hprof}
{HPROF Parser}.
\newblock \url{https://github.com/eaftan/hprof-parser}.

\bibitem{jensen}
C.~S. Jensen, M.~R. Prasad, and A.~Moller.
\newblock Automated testing with targeted event sequence generation.
\newblock In {\em ISSTA}, 2013.

\bibitem{lazy}
S.~Khurshid, C.~S. P\u{a}s\u{a}reanu, and W.~Visser.
\newblock Generalized symbolic execution for model checking and testing.
\newblock In {\em TACAS}, 2003.

\bibitem{miller}
B.~P. Miller, G.~Cooksey, and F.~Moore.
\newblock An empirical study of the robustness of macos applications using
  random testing.
\newblock In {\em Proceedings of the International Workshop on Random Testing},
  2006.

\bibitem{sigdroid}
N.~Mirzaei, H.~Bagheri, R.~Mahmood, and S.~Malek.
\newblock {SIG-Droid}: automated system input generation for android
  applications.
\newblock In {\em ISSRE}, 2015.

\bibitem{mulliner2009injecting}
C.~Mulliner and C.~Miller.
\newblock Injecting sms messages into smart phones for security analysis.
\newblock In {\em WOOT}, 2009.

\bibitem{combined-spf}
C.~S. P\u{a}s\u{a}reanu, P.~C. Mehlitz, D.~H. Bushnell, K.~Gundy-Burlet,
  M.~Lowry, S.~Person, and M.~Pape.
\newblock Combining unit-level symbolic execution and system-level concrete
  execution for testing nasa software.
\newblock In {\em ISSTA}, 2008.

\bibitem{spf}
C.~S. P\u{a}s\u{a}reanu, W.~Visser, D.~Bushnell, J.~Geldenhuys, P.~Mehlitz, and
  N.~Rungta.
\newblock {Symbolic PathFinder}: integrating symbolic execution with model
  checking for java bytecode analysis.
\newblock In {\em ASE}, 2013.

\bibitem{ucklee}
D.~A. Ramos and D.~Engler.
\newblock {Under-Constrained Symbolic Execution}: correctness checking for real
  code.
\newblock In {\em USENIX Security}, 2015.

\bibitem{under2}
D.~A. Ramos and D.~R. Engler.
\newblock Practical, low-effort equivalence verification of real code.
\newblock In {\em CAV}, 2011.

\bibitem{task}
C.~Ren, Y.~Zhang, H.~Xue, T.~Wei, and P.~Liu.
\newblock Towards discovering and understanding task hijacking in android.
\newblock In {\em USENIX Security}, 2015.

\bibitem{all-taint}
E.~J. Schwartz, T.~Avgerinos, and D.~Brumley.
\newblock All you ever wanted to know about dynamic taint analysis and forward
  symbolic execution (but might have been afraid to ask).
\newblock In {\em S\&P}, 2010.

\bibitem{jpf-nhandler}
N.~Shafiei and F.~van Breugel.
\newblock Automatic handling of native methods in {Java PathFinder}.
\newblock In {\em SPIN Symposium on Model Checking of Software}, 2014.

\bibitem{kratos}
Y.~Shao, J.~Ott, Q.~A. Chen, Z.~Qian, and Z.~M. Mao.
\newblock {Kratos}: discovering inconsistent security policy enforcement in the
  android framework.
\newblock In {\em NDSS}, 2016.

\bibitem{stagefright}
Stagefright.
\newblock \url{https://en.wikipedia.org/wiki/Stagefright_(bug)}.

\bibitem{jpf}
W.~Visser, K.~Havelund, G.~Brat, S.~Park, and F.~Lerda.
\newblock Model checking programs.
\newblock In {\em ASE}, 2003.

\bibitem{users}
WSJ.
\newblock {Google says android has 1.4 billion active users}.
\newblock
  \url{www.wsj.com/articles/google-says-android-has-1-4-billion-active-users-1443546856}.

\end{thebibliography}
%
%

\end{document}